\documentclass[aps,prl,reprint,superscriptaddress]{revtex4-2}
\usepackage{amsmath,bm,mathrsfs,color}
\usepackage{graphicx}

\begin{document}

% Use the \preprint command to place your local institutional report
% number in the upper righthand corner of the title page in preprint mode.
% Multiple \preprint commands are allowed.
% Use the 'preprintnumbers' class option to override journal defaults
% to display numbers if necessary
%\preprint{}

%Title of paper
\title{%\underline{\small{\tt ver. 0.6 (\today)}}
%\vspace{3mm}\\
Topological blocking at the Bi(111) surface due to surface relaxation
%Topological blocking by surface relaxation: Concealment of surface states in Bi(111)
}

% repeat the \author .. \affiliation  etc. as needed
% \email, \thanks, \homepage, \altaffiliation all apply to the current
% author. Explanatory text should go in the []'s, actual e-mail
% address or url should go in the {}'s for \email and \homepage.
% Please use the appropriate macro foreach each type of information

% \affiliation command applies to all authors since the last
% \affiliation command. The \affiliation command should follow the
% other information
% \affiliation can be followed by \email, \homepage, \thanks as well.
\author{Kazuki Koie}
\affiliation{Department of Engineering Science, University of Electro-Communications, Tokyo 182-8585, Japan}
\affiliation{Department of Physics, Kobe University, Kobe, Japan}
\author{Rikako Yaguchi}
\affiliation{Department of Engineering Science, University of Electro-Communications, Tokyo 182-8585, Japan}
\author{Yuki Fuseya}
\affiliation{Department of Engineering Science, University of Electro-Communications, Tokyo 182-8585, Japan}
\affiliation{Department of Physics, Kobe University, Kobe 657-8501, Japan}
%\email[]{Your e-mail address}
%\homepage[]{Your web page}
%\thanks{}

%Collaboration name if desired (requires use of superscriptaddress
%option in \documentclass). \noaffiliation is required (may also be
%used with the \author command).
%\collaboration can be followed by \email, \homepage, \thanks as well.
%\collaboration{}
%\noaffiliation

\date{\today}

\begin{abstract}
The topological characteristics of Bi and its alloys with Sb have fueled intense debate since the prediction of three-dimensional topological insulators. However, a definitive resolution has not been reached to date. Here, we provide theoretical evidence that surface relaxation conceals the underlying bulk topology of pure Bi. 
Using density functional theory calculations for thin Bi(111) films (up to 17 bilayers), we first demonstrate a substantial inter-bilayer expansion near the surface. Motivated by this finding, we extend our analysis to thick Bi(111) films (up to 250 bilayers) incorporating relaxation layers, within the framework of a relativistic empirical tight-binding model. Our results reveal that these relaxation layers topologically block the emergence of surface state and significantly suppress the one-particle spectrum of surface states, thereby obscuring the experimental identification of Bi's topological properties. 
This phenomenon, which we term ``topological blocking", provides crucial insights into the long-standing difficulty of observing surface states of Bi(111) at the $\bar{M}$ point. Furthermore, it establishes a framework for understanding and predicting the topological behavior in systems where surface relaxation disrupts the bulk--edge correspondence.

\end{abstract}

% insert suggested keywords - APS authors don't need to do this
%\keywords{}

%\maketitle must follow title, authors, abstract, and keywords
\maketitle

% body of paper here - Use proper section commands
% References should be done using the \cite, \ref, and \label commands
%\section{Introduction}
Since its prediction as a topological insulator, bismuth (Bi) alloyed with antimony (Sb) has gained attention in the study of topological materials \cite{Fu2007}. The surface states of Bi and BiSb have been intensively explored both experimentally \cite{Hsieh2008,Hsieh2009,Hirahara2010,Nishide2010,HGuo2011,Nakamura2011,Ohtsubo2013,Benia2015,Ito2016,Ohtsubo2016,Fukushima2023} and theoretically \cite{Teo2008,HJZhang2009,Aguilera2015,Fuseya2018,Aguilera2021}, yet their precise topological nature remains unclear.
A fundamental question centers around the surface states at the $\bar{M}$ point in the (111)-surface Brillouin zone (corresponds to the $L$ point in the bulk Brillouin zone), with profound implications for topological characterization.

Theoretically, various studies have consistently reached the same conclusions \cite{Ferreira1967,Ferreira1968,Golin1968,Gonze1988,Gonze1990,Shick1999,Timrov2012,Aguilera2015}. 
For pure bulk Bi, the conduction band at the $L$ point is symmetric, whereas the valence band is antisymmetric, indicating a \emph{trivial} $Z_2$ topological invariant \cite{Fu2007}. (We denote this band order as $L_s/L_a$.) All density-functional-theory (DFT) support this band order. 
When pure Bi is topologically trivial, the two surface states (S1 and S2) cross linearly (i.e., zero-gap Dirac-like surface state) at the $\bar{M}$ point, as confirmed by numerical \cite{Fu2007,Teo2008} and analytical \cite{Fuseya2018} studies.
(It is worth noting that the existence condition for topological surface states at the $\bar{M}$ point in Bi is opposite to that of ordinary topological insulators, where the nontrivial topology is manifested by the presence of Dirac-like surface states \cite{Fuseya2018,Aguilera2021}.)
In contrast, if Bi is topologically nontrivial, a finite gap opens between the two surface states.
Alternatively, gap opening can occur by the interference between opposite surface states \cite{BZhou2008,Linder2009,HZLu2010,SQShen_book,Ozawa2014}. According to such surface interference scenario, a sizable surface gap can appear even when Bi is topologically trivial \cite{Fuseya2018,Aguilera2021}, suggesting a surface-sensitive topological nature of Bi.

Experimentally, however, a clear division persists. Angle-resolved photoemission spectroscopy (ARPES) measurements often indicate a \emph{nontrivial} topology of Bi, as evidenced by a surface gap at the $\bar{M}$ point \cite{Ohtsubo2013,Ito2016,Ohtsubo2016,Fukushima2023}, except for a few reports (e.g., \cite{Benia2015}). In contrast, scanning tunneling microscopy and transport studies suggest a \emph{trivial} topology supported by a signature for the higher-order topology in Bi \cite{Schindler2018b,Aggarwal2021}, which is nothing but evidence for the trivial $Z_2$ topology. Thus, the evidence for the higher-order topology deepens the discrepancy in interpretation. Hereafter, we use the term ``trivial" or ``nontrivial" to indicate the lowest-order topology characterized by the $Z_2$ topological invariance.

A key reason for the abovementioned discrepancy may be attributed to the remarkable sensitivity of the topology of Bi to the lattice constant.
Only a change of approximately $0.4$\% in the lattice constant is sufficient to invert the bands at the $L$ point, triggering a topological transition from trivial to nontrivial \cite{Aguilera2015}.
For example, a strain-induced topological transition has been demonstrated in ultrathin Bi(111) on Bi$_2$Te$_3$(111) substrate \cite{Hirahara2012}.
Furthermore, temperature-driven topological transitions owing to thermal lattice distortion have been shown in ARPES measurements \cite{Ohtsubo2019}.

In this study, we revealed a pronounced expansion of inter-bilayer distance near the surface of Bi(111), and such an expanded surface layer topologically prevents surface states from appearing.
These insights into the elusive topological characteristics of Bi represent a critical step toward resolving the outstanding debate mentioned above.

%%%%%%%%%%%%%%%%%%%%%%%%%%%%%%%%%%%%%%%%%%%%%%%%%%%%%%%%%%%%
%\section{Theory}
%\subsection{inter-bilayer expansion near the surface}
%======================================================================== 
 \begin{figure}[tb]
 \includegraphics[width=8cm]{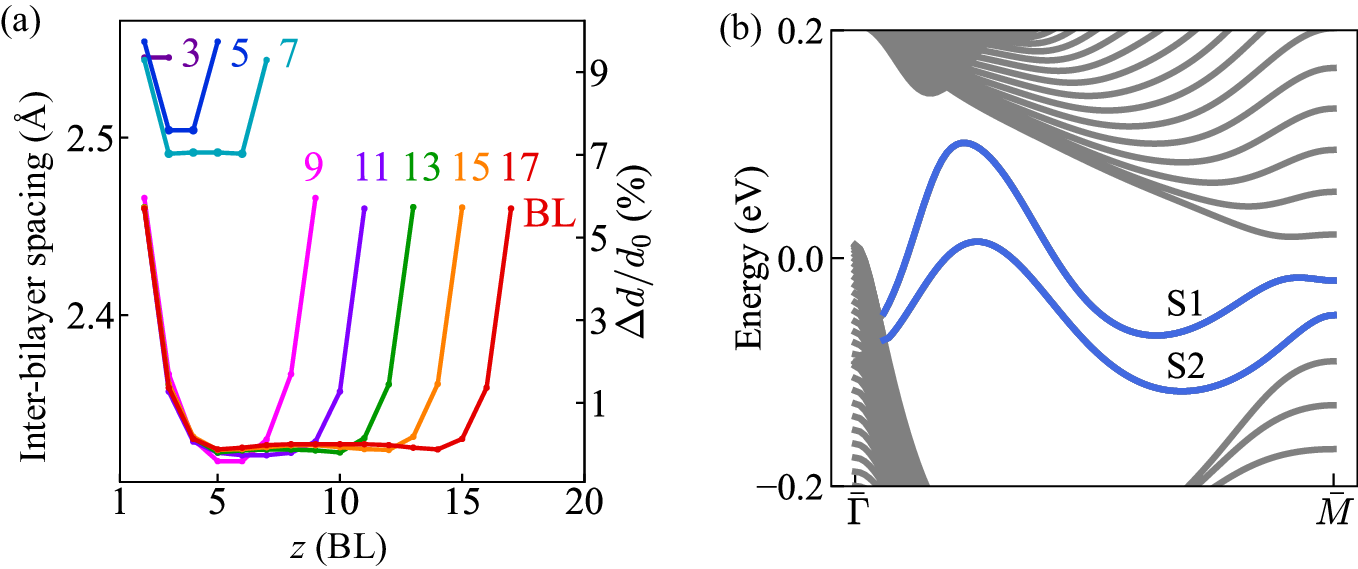}
 \caption{\label{Fig1}	(a) Inter-bilayer spacing of Bi(111) film for 3--17 bilayer (BL) as a function of the distance $z$ from the surface computed using DFT with structural optimization. The inter-bilayer spacing of bulk Bi is 2.35 \AA \cite{Monig2005,Hofmann2006}.
	(b) Energy eigenvalues of 100-BL Bi(111) based on the Liu--Allen model with surface potential \cite{Saito2016,Asaka2022} for $\Delta d=0$.
}
 \end{figure}
%========================================================================

In general, the interlayer distance can be changed by surface relaxation \cite{Davis1992}. To examine how surface relaxation occurs in Bi, we investigated the lattice displacement in Bi(111) films using DFT with the BAND software from the Amsterdam Modeling Suite \cite{Velde1991,BAND}. Structural optimization was performed using a fast inertial relaxation engine \cite{Bitzek2006} with the Perdew--Burke--Ernzerhof exchange-correlation functional from LibXC \cite{Perdew1996}, double-zeta-polarized basis sets, and numerical orbitals with a small frozen core. Relativistic effects were incorporated using the zeroth-order regular approximation (ZORA) \cite{Lenthe1999}.
Figure \ref{Fig1}(a) shows inter-bilayer spacing $d$ as a function of the distance $z$ from the surface for various film thicknesses (3--17 bilayers---BL), incorporating the ZORA scalar relativistic effects. (We have confirmed that for thin bilayers, the results obtained with scalar relativistic effects qualitatively agree with those obtained using fully relativistic spin-orbit coupling.)
Near the surface, inter-bilayer spacing expands by up to 6\% relative to central layer spacing $d_0$, averaging $\Delta d / d_0 \sim3$\% across approximately five surface bilayers, while the atoms remain fixed along the in-plane direction. Such a significant inter-bilayer expansion for Bi(111) films is consistent with the previous reports \cite{Monig2005,Koroteev2008,Hirahara2012}.
Figure \ref{Fig1}(a) clearly demonstrates that the inter-bilayer expansion behavior converges for thicknesses exceeding 11 BL. This convergence is reasonable, as the relaxation layer has a thickness of 5 BL. Consequently, the properties of the relaxation layer should remain robust in much thicker films.

Prior to discussing the main findings of this work, we briefly outline the surface states of Bi(111). Figure \ref{Fig1}(b) presents the eigenvalues for a 100-BL Bi(111) film without inter-bilayer modulation, calculated based on the Liu--Allen relativistic empirical tight-binding model \cite{Liu1995,Saito2016,Fuseya2018,Asaka2022}, exhibiting exceptional agreement with experimental results for Bi. We employed the surface potential introduced by Saito {\it et al.} \cite{Saito2016,Asaka2022}, being consistent with ARPES measurements \cite{Ast2003,Koroteev2004,Hirahara2006,Ohtsubo2013,Benia2015,Ito2016}.
Consistent with other DFT results \cite{Ferreira1967,Ferreira1968,Golin1968,Gonze1988,Gonze1990,Shick1999,Timrov2012,Aguilera2015}, the Liu--Allen model identifies pure Bi as topologically \emph{trivial}, characterized by the $L_s/L_a$ configuration. 
The surface states S1 and S2 merge into the valence band at the $\bar{\Gamma}$ point and reside within the bulk bandgap at the $\bar{M}$ point. 
The overall dispersion of the surface states remains qualitatively unchanged even when the bulk band at the $L$-point is inverted due to the inter-bilayer expansion, suggesting that inter-bilayer expansion near the surface does not lead to an immediately discernible modification of the surface states.
However, a profound alteration, obscured in the eigenvalue plot, emerges upon examining the wavefunction characteristics or the one-particle spectral function.

\begin{figure}[tb]
    \includegraphics[width=8cm]{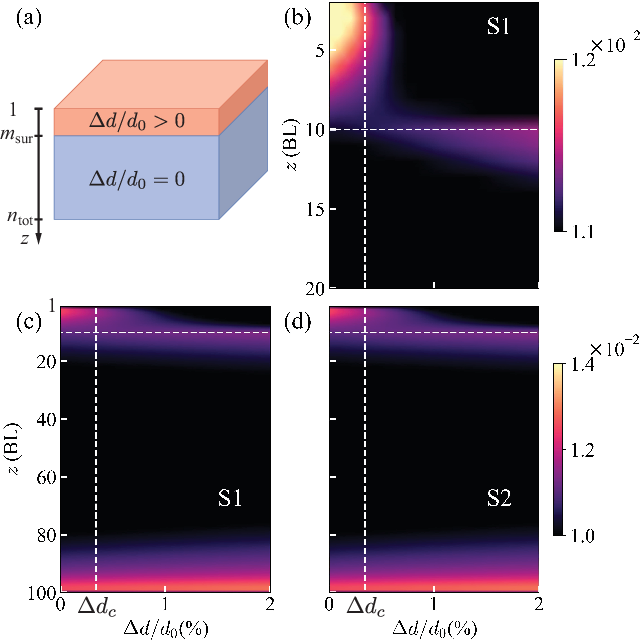}
    \caption{\label{Fig4}
    (a) Illustration of film considering surface relaxation with expanded inter-bilayer spacing ($\Delta d/d_0 >0$) for $z \le m_{\rm sur}$.
    (b) Probability distribution $|\psi(z)|^2$ with surface relaxation ($n_{\rm tot}=100$ and $m_{\rm sur}=10$) near the surface ($z \le 20$ BL) of surface state S1. 
    Entire range plot $1\le z \le 100$ BL is given for surface states (c) S1 and (d) S2.
    }
\end{figure}

%\emph{inter-bilayer expansion near the surface.}
To examine the effects of surface relaxation, we considered a Bi(111) film with a total thickness of $n_{\rm tot}$ BL, where the inter-bilayer distance was relaxed for $m_{\rm sur}$ BL on one side of the surface, and the remaining inter-bilayer distance is unchanged from the bulk value, as depicted in Fig. \ref{Fig4}(a).
%\subsubsection{Wavefunction}
%========================================================================
%========================================================================
%\emph{Wave function.}
Figure \ref{Fig4} (b)-(d) shows $\Delta d$-dependences of probability distribution $|\psi(z)|^2$ of $n_{\rm tot}=100$ with $m_{\rm sur}=10$ for surface states S1 [Fig. \ref{Fig4}(c)] and S2 [Fig. \ref{Fig4}(d)], where $d_0$ is the inter-bilayer spacing for bulk used in the Liu--Allen model. (Using the general relationship between interatomic matrix element $V$ and its distance $V\propto d^{-2}$ \cite{Liu1995}, we estimated the changes in $\Delta d$ using $\Delta d /d_0 =\left(1+\Delta V/V_0 \right)^{-1/2}-1$.)
In addition, $|\psi(z)|^2$ for S1 near the surface with $z\le 20$ BL is shown in Fig. \ref{Fig4} (b).
There is a characteristic value for the inter-bilayer spacing, $\Delta d_c$, at which the spatial distribution of $|\psi(z)|^2$ undergoes a drastic transformation, indicated by the vertical dashed lines. (The origin of $\Delta d_c$ will be elucidated in a later discussion, where $\Delta d_c/d_0 = 0.34$ \% will be determined.)
For $\Delta d \lesssim \Delta d_c$, the wavefunction exists on both surfaces, $z=1$ and 100 BL. In contrast, when $\Delta d \gtrsim \Delta d_c$, the wavefunction shifts beneath the surface layer, demonstrating that the surface states effectively migrate below the relaxation layer. 
Thus, the relaxation layer acts as a barrier that prevents the surface state from appearing, effectively blocking its manifestation at the surface.

%\subsubsection{One-particle spectrum}
%========================================================================
\begin{figure}
    \includegraphics[width=8cm]{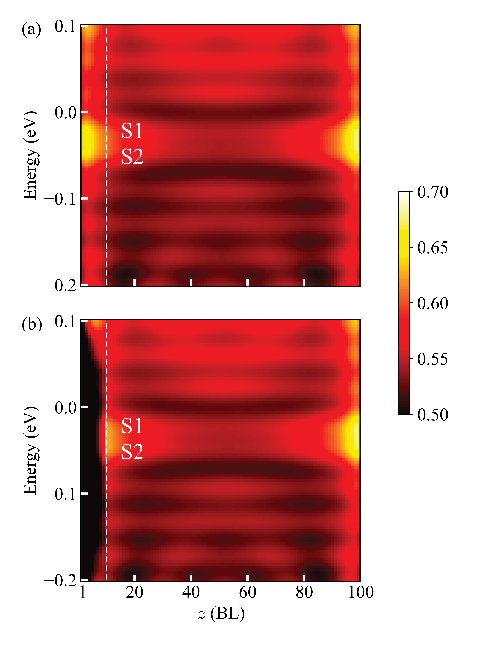}
    \caption{\label{Fig5} Spatial one-particle spectrum $A(\bm{k}_\parallel, z, \varepsilon)$ of 100-BL Bi(111) with surface relaxation ($n_{\rm tot}=100$ and $m_{\rm sur}=10$) for (a) $\Delta d /d_0 = 0.2$\% (surface layers are trivial) and (b) $\Delta d /d_0 = 3$\% (surface layers are nontrivial).
    }
\end{figure}
%========================================================================

%\emph{One-particle spectrum.}
The following spatial one-particle spectral function, $A(\bm{k}_\parallel, z, \varepsilon )$, can provide further insights into this blocking effect:
\begin{align}
	A(\bm{k}_\parallel, z, \varepsilon ) &= -\frac{1}{\pi} {\rm Tr}_z {\rm Im} G^R (\bm{k}_\parallel, z, \varepsilon ),
\end{align}
where $
	G^R(\bm{k}_\parallel, z, \varepsilon ) =\left[ \varepsilon - \mathcal{H} (\bm{k}_\parallel, z) +i\varSigma'' \right]^{-1} 
$ is the retarded Green function, $\varSigma''$ denotes the imaginary part of the self-energy, and Tr$_z$ is the trace over $G^R$ for a particular $z$-th BL of interest \cite{Asaka2022}.
The spatial one-particle spectrum at the $\bar{M}$ point is shown in Fig. \ref{Fig5} for $\Delta d/d_0 = 0.2$ \% [Fig. \ref{Fig5}(a)] and $\Delta d/d_0=3$ \% [Fig. \ref{Fig5}(b)] with $\varSigma''=0.03$ eV. 
For $\Delta d /d_0= 0.2$ \% ($<\Delta d_c/d_0$), surface states S1 and S2 are observed on both surfaces at $z=1$ and 100 BL.
By contrast, for $\Delta d /d_0= 3$ \% ($>\Delta d_c/d_0$), S1 and S2 exist on one side of the surface ($z=100$ BL) but not on the other side ($z=1$ BL). 
This result clearly indicates that the relaxation layer of $z=1$--10 BL blocks the two surface states only for $\Delta d/d_0 = 3$ \%, and does not for $\Delta d/d_0 = 0.2$ \%.
The absence of a surface state blocked by the relaxation layer results in the absence of photoemission intensity.

%\subsection{Thickness dependence}
%========================================================================
\begin{figure}
    \includegraphics[width=8cm]{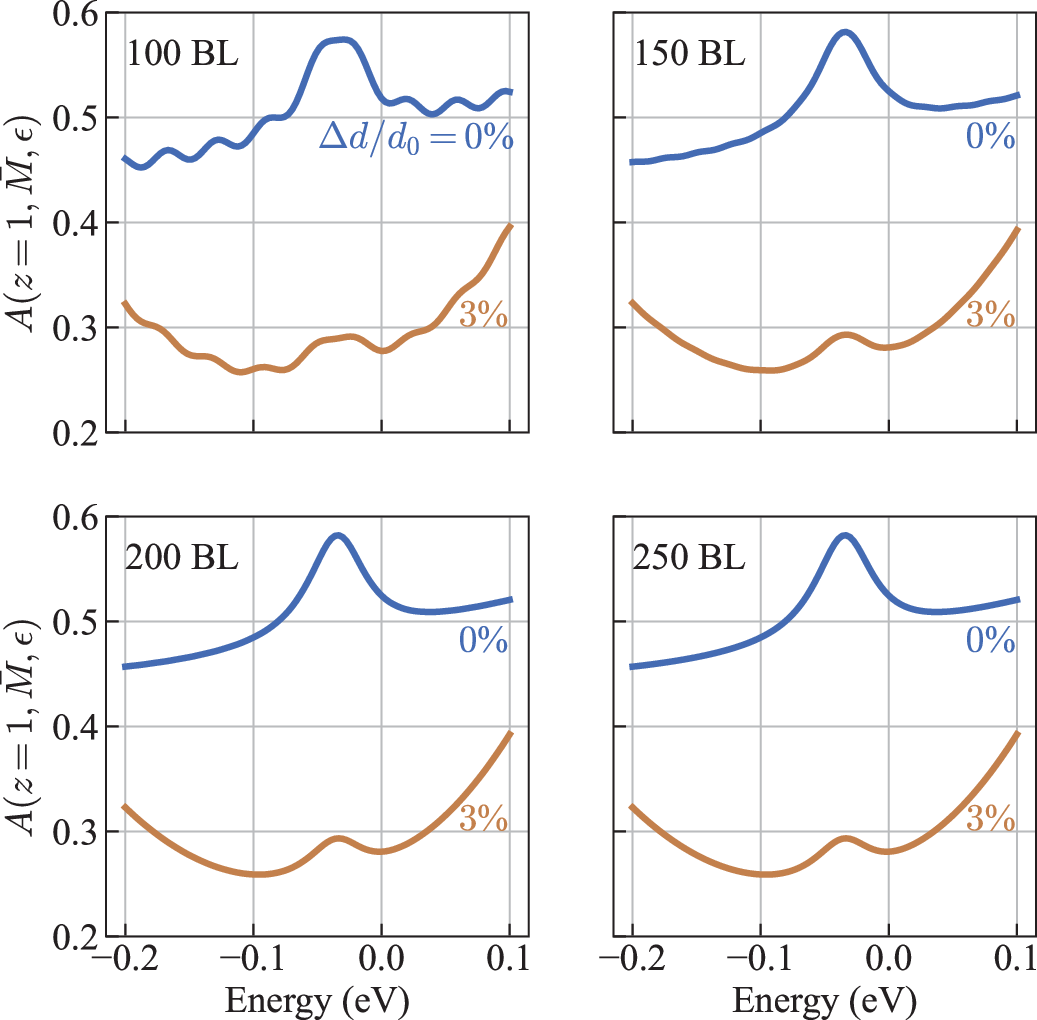}
    \caption{\label{Fig6}
    One-particle spectra at $\bar{M}$ point with and without surface expansion. The total thickness, $n_{\rm tot}$, varies from $100$ to $250$, whereas the thickness of the surface relaxation layer is set to $m_{\rm sur}=10 $. The blocking ratio is 0.49 for a 100-BL film and 0.50 for a 150--250-BL film.
    }
\end{figure}
%========================================================================

%\emph{Thickness dependence.}
A comparison of one-particle spectra at the surface ($z=1$ BL) between relaxed ($\Delta d/d_0 =3$\%) and unrelaxed ($\Delta d=0$) surface layers is shown in Fig. \ref{Fig6}  for thickness $n_{\rm tot}$ varying from $100$ to $250$ and keeping $m_{\rm sur}=10$. The impact of the blocking can be estimated using the blocking ratio, which is the ratio of one-particle spectra with and without relaxed layers.
The blocking ratio of the peak intensity is 0.49 for a 100 BL film and 0.50 for 150--250-BL films, indicating that the effect is independent of the film thickness. Therefore, the blocking persists even in thick Bi(111) slabs.
Although the peak position remains detectable with $m_{\rm sur}=10$, the intensity of the surface states is substantially weaker than that of the bulk state. Consequently, the blocking hinders the detection of the surface states of Bi(111), masking evidence of the topological nature of Bi.

%========================================================================
\begin{figure}
    \includegraphics[width=3.8cm]{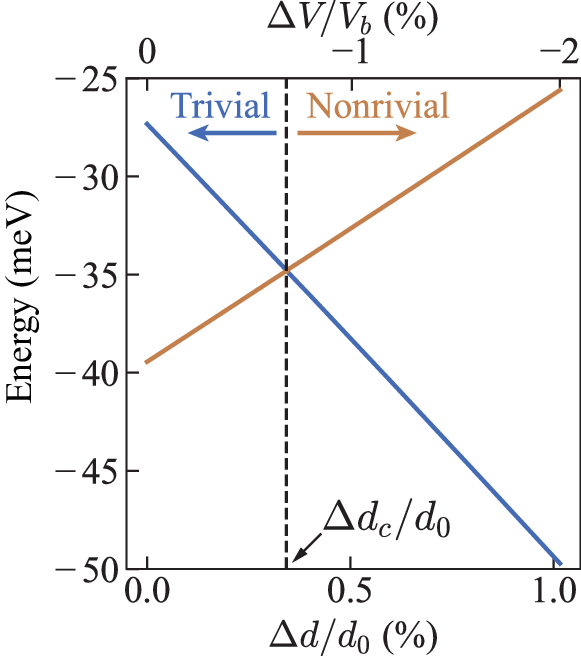}
    \caption{\label{Fig7}
    Energy shift at $L$ point of \emph{bulk} Bi as a function of inter-bilayer expansion $\Delta d=d-d_0$ using the Liu--Allen model \cite{Liu1995}. 
    }
\end{figure}
%========================================================================

Let us now elucidate why the relaxation layer obstructs the emergence of surface states only for $\Delta d \gtrsim \Delta d_c$.
Figure \ref{Fig7} presents the energy shift of the conduction and valence bands at the $L$ point of \emph{bulk} Bi as a function of $\Delta d$. 
As $\Delta d$ increases, $|E_g|$ decreases linearly, leading to a band inversion at $\Delta d_c / d_0 = 0.34$\% ($\Delta V_c/ V_0 = -0.68$\%). This band inversion drives a transition in the three-dimensional $Z_2$ topological invariants $(\nu_0; \nu_1 \nu_2 \nu_3)$ from $(0;000)$ to $(1;111)$, signifying a transition from a trivial to a nontrivial topology \cite{Fu2007,Teo2008}. 
In general, such a toplogical transition becomes less evident as the film thickness decreases, at least when considering the energy profile of the surface states \cite{Fuseya2018,Aguilera2021}. However, its signature persists as a crossover in the wave function distribution $|\psi(z)|^2$ (Fig. \ref{Fig4}). Remarkably, this crossover in the relaxation layer (white dashed lines in Fig. \ref{Fig4}) occurs at nearly the same position as the bulk band inversion point $\Delta d_c /d_0 = 0.34$ \%. 
This striking agreement strongly suggests that the qualitative change in $|\psi(z)|^2$, and the resultant blocking effect, originates from the topological transition.

For the trivial $L_s/L_a$ configuration, it has been analytically shown that two Dirac-like surface states must exist within the bandgap at the $\bar{M}$ point \cite{Fuseya2018}.
In contrast, for the nontrivial $L_a/L_s$ configuration, surface states are strictly prohibited at the $\bar{M}$ point.
This prohibition underlies the observed blocking effect, whereby the emergence of surface states is suppressed when $\Delta d \gtrsim \Delta d_c$.

Conseqently, the blocking effect is inherently rooted in the topological properties of Bi. For $\Delta d \gtrsim \Delta d_c$, the relaxation layer effectively acquires the characteristics of a topologically nontrivial state, leading to the supression of surface states. This mechanism constitutes the essence of the ``topolgical blocking effect".

While our results mainly consider $m_{\rm sur}=10$, we checked that the topological blocking persists for thinner relaxation layers, such as $m_{\rm sur}=5$. As inter-bilayer expansion $\Delta d$ varies as a function of $z$, as shown in Fig. \ref{Fig1}(a), the uniform $\Delta d$ values in our analysis may slightly overestimate topological blocking. However, experimental observations report an inter-bilayer expansion of approximately $\Delta d/d_0 =$ 2--3\% \cite{Monig2005,Hirahara2012}, which is substantially larger than critical threshold $\Delta d_c /d_c = 0.34$\% and is enough to induce the topological transition. Thus, the observed topological blocking mechanism is highly plausible and relevant to real systems.

Owing to restrictions on computational resources, we combined DFT for structural optimization and the tight-binding method for obtaining the probability distribution and spatial one-particle spectrum.
The DFT-based computation of $A(\bm{k}_\parallel, z, \varepsilon)$ with structural optimization may allow to verify topological blocking directly. This verification will be explored in future work.

%\section{Summary}

In conclusion, we have unveiled a novel mechanism that can obscure a material's topological nature: the topological blocking effect. 

The key findings of this study are as follows:
(i) A pronounced inter-bilayer expansion was observed near the surface of Bi(111) films.
(ii) The relaxation layer effectively blocked the emergence of surface states and significantly suppressed the one-particle spectrum of surface states in Bi(111). 
(iii) The blocking effect occurs only for $\Delta d \gtrsim \Delta d_c$, where the boundary value of $\Delta d_c$ exhibits remarkable agreement with the topological transition point for the bulk Bi.
Based on these findings, we conclude that the blocking effect originates from the nontrivial topological nature in the relaxation layer.
Thus, even if the bulk of material possesses the topologically trivial character, its topological signature at the surface can be masked by the topological blocking effect within the relaxation layer. This blocking effect persisted even in thick slabs, complicating the experimental identification of topologically trivial nature in Bi. This results in a spontaneous breaking of the bulk--edge correspondence.

Our proposal that surface relaxation can lead to the spontaneous breaking of the bulk--edge correspondence is not limited to Bi but can be broadly applied to other systems. Conventional bulk--edge correspondence assumes that the crystal structure near the surface remains identical to that of the bulk. However, it is well-established that the surface undergoes structural modifications, such as surface relaxation and reconstruction, deviating from the bulk crystal structure \cite{Davis1992,Oura_book}. This fundamental mismatch challenges the validity of the bulk--edge correspondence in real systems.
Surface relaxation does not universally disrupt the bulk--edge correspondence. In some cases, surface relaxation may leave the band structure qualitatively unchanged. However, in many topological materials, such as Sb, PbTe, and SnTe \cite{Hsieh2012}, the bulk band structure is poised near a topological transition. In such systems, even modest surface relaxation can induce a qualitative change in the electronic structure, thereby breaking the bulk--edge correspondence.
Bismuth is an exemplary material for observing this phenomenon because of its small band gap and large inter-bilayer expansion. 

The abovementioned insights can lead to establish a crucial framework for interpreting experimental data where the bulk--edge correspondence alone may be insufficient, thus paving the way for a deeper understanding of the topological properties of materials with surface relaxation effects.

%\section{Data availability statement}

% If you have acknowledgments, this puts in the proper section head.
\begin{acknowledgments}
We thank Y. Asaka for the fruitful discussions.
This work was supported by the Japan Society for the Promotion of Science [Grants No. 23H00268, 23H04862 and 22K18318].
\end{acknowledgments}

%\section{Author contributions}

% Create the reference section using BibTeX:
\bibliography{Bismuth,footnote}

%apsrev4-2.bst 2019-01-14 (MD) hand-edited version of apsrev4-1.bst
%Control: key (0)
%Control: author (8) initials jnrlst
%Control: editor formatted (1) identically to author
%Control: production of article title (0) allowed
%Control: page (0) single
%Control: year (1) truncated
%Control: production of eprint (0) enabled
\begin{thebibliography}{50}%
\makeatletter
\providecommand \@ifxundefined [1]{%
 \@ifx{#1\undefined}
}%
\providecommand \@ifnum [1]{%
 \ifnum #1\expandafter \@firstoftwo
 \else \expandafter \@secondoftwo
 \fi
}%
\providecommand \@ifx [1]{%
 \ifx #1\expandafter \@firstoftwo
 \else \expandafter \@secondoftwo
 \fi
}%
\providecommand \natexlab [1]{#1}%
\providecommand \enquote  [1]{``#1''}%
\providecommand \bibnamefont  [1]{#1}%
\providecommand \bibfnamefont [1]{#1}%
\providecommand \citenamefont [1]{#1}%
\providecommand \href@noop [0]{\@secondoftwo}%
\providecommand \href [0]{\begingroup \@sanitize@url \@href}%
\providecommand \@href[1]{\@@startlink{#1}\@@href}%
\providecommand \@@href[1]{\endgroup#1\@@endlink}%
\providecommand \@sanitize@url [0]{\catcode `\\12\catcode `\$12\catcode
  `\&12\catcode `\#12\catcode `\^12\catcode `\_12\catcode `\%12\relax}%
\providecommand \@@startlink[1]{}%
\providecommand \@@endlink[0]{}%
\providecommand \url  [0]{\begingroup\@sanitize@url \@url }%
\providecommand \@url [1]{\endgroup\@href {#1}{\urlprefix }}%
\providecommand \urlprefix  [0]{URL }%
\providecommand \Eprint [0]{\href }%
\providecommand \doibase [0]{https://doi.org/}%
\providecommand \selectlanguage [0]{\@gobble}%
\providecommand \bibinfo  [0]{\@secondoftwo}%
\providecommand \bibfield  [0]{\@secondoftwo}%
\providecommand \translation [1]{[#1]}%
\providecommand \BibitemOpen [0]{}%
\providecommand \bibitemStop [0]{}%
\providecommand \bibitemNoStop [0]{.\EOS\space}%
\providecommand \EOS [0]{\spacefactor3000\relax}%
\providecommand \BibitemShut  [1]{\csname bibitem#1\endcsname}%
\let\auto@bib@innerbib\@empty
%</preamble>
\bibitem [{\citenamefont {Fu}\ and\ \citenamefont {Kane}(2007)}]{Fu2007}%
  \BibitemOpen
  \bibfield  {author} {\bibinfo {author} {\bibfnamefont {L.}~\bibnamefont
  {Fu}}\ and\ \bibinfo {author} {\bibfnamefont {C.~L.}\ \bibnamefont {Kane}},\
  }\bibfield  {title} {\bibinfo {title} {Topological insulators with inversion
  symmetry},\ }\href@noop {} {\bibfield  {journal} {\bibinfo  {journal} {Phys.
  Rev. B}\ }\textbf {\bibinfo {volume} {76}},\ \bibinfo {pages} {045302}
  (\bibinfo {year} {2007})}\BibitemShut {NoStop}%
\bibitem [{\citenamefont {Hsieh}\ \emph {et~al.}(2008)\citenamefont {Hsieh},
  \citenamefont {Qian}, \citenamefont {Wray}, \citenamefont {Xia},
  \citenamefont {Hor}, \citenamefont {Cava},\ and\ \citenamefont
  {Hasan}}]{Hsieh2008}%
  \BibitemOpen
  \bibfield  {author} {\bibinfo {author} {\bibfnamefont {D.}~\bibnamefont
  {Hsieh}}, \bibinfo {author} {\bibfnamefont {D.}~\bibnamefont {Qian}},
  \bibinfo {author} {\bibfnamefont {L.}~\bibnamefont {Wray}}, \bibinfo {author}
  {\bibfnamefont {Y.}~\bibnamefont {Xia}}, \bibinfo {author} {\bibfnamefont
  {Y.~S.}\ \bibnamefont {Hor}}, \bibinfo {author} {\bibfnamefont {R.~J.}\
  \bibnamefont {Cava}},\ and\ \bibinfo {author} {\bibfnamefont {M.~Z.}\
  \bibnamefont {Hasan}},\ }\bibfield  {title} {\bibinfo {title} {A topological
  dirac insulator in a quantum spin hall phase},\ }\href@noop {} {\bibfield
  {journal} {\bibinfo  {journal} {Nature}\ }\textbf {\bibinfo {volume} {452}},\
  \bibinfo {pages} {970} (\bibinfo {year} {2008})}\BibitemShut {NoStop}%
\bibitem [{\citenamefont {Hsieh}\ \emph {et~al.}(2009)\citenamefont {Hsieh},
  \citenamefont {Xia}, \citenamefont {Wray}, \citenamefont {Qian},
  \citenamefont {Pal}, \citenamefont {Dil}, \citenamefont {Osterwalder},
  \citenamefont {Meier}, \citenamefont {Bihlmayer}, \citenamefont {Kane},
  \citenamefont {Hor}, \citenamefont {Cava},\ and\ \citenamefont
  {Hasan}}]{Hsieh2009}%
  \BibitemOpen
  \bibfield  {author} {\bibinfo {author} {\bibfnamefont {D.}~\bibnamefont
  {Hsieh}}, \bibinfo {author} {\bibfnamefont {Y.}~\bibnamefont {Xia}}, \bibinfo
  {author} {\bibfnamefont {L.}~\bibnamefont {Wray}}, \bibinfo {author}
  {\bibfnamefont {D.}~\bibnamefont {Qian}}, \bibinfo {author} {\bibfnamefont
  {A.}~\bibnamefont {Pal}}, \bibinfo {author} {\bibfnamefont {J.~H.}\
  \bibnamefont {Dil}}, \bibinfo {author} {\bibfnamefont {J.}~\bibnamefont
  {Osterwalder}}, \bibinfo {author} {\bibfnamefont {F.}~\bibnamefont {Meier}},
  \bibinfo {author} {\bibfnamefont {G.}~\bibnamefont {Bihlmayer}}, \bibinfo
  {author} {\bibfnamefont {C.~L.}\ \bibnamefont {Kane}}, \bibinfo {author}
  {\bibfnamefont {Y.~S.}\ \bibnamefont {Hor}}, \bibinfo {author} {\bibfnamefont
  {R.~J.}\ \bibnamefont {Cava}},\ and\ \bibinfo {author} {\bibfnamefont
  {M.~Z.}\ \bibnamefont {Hasan}},\ }\bibfield  {title} {\bibinfo {title}
  {Observation of unconventional quantum spin textures in topological
  insulators},\ }\href@noop {} {\bibfield  {journal} {\bibinfo  {journal}
  {Science}\ }\textbf {\bibinfo {volume} {323}},\ \bibinfo {pages} {919}
  (\bibinfo {year} {2009})}\BibitemShut {NoStop}%
\bibitem [{\citenamefont {Hirahara}\ \emph {et~al.}(2010)\citenamefont
  {Hirahara}, \citenamefont {Sakamoto}, \citenamefont {Saisyu}, \citenamefont
  {Miyazaki}, \citenamefont {Kimura}, \citenamefont {Okuda}, \citenamefont
  {Matsuda}, \citenamefont {Murakami},\ and\ \citenamefont
  {Hasegawa}}]{Hirahara2010}%
  \BibitemOpen
  \bibfield  {author} {\bibinfo {author} {\bibfnamefont {T.}~\bibnamefont
  {Hirahara}}, \bibinfo {author} {\bibfnamefont {Y.}~\bibnamefont {Sakamoto}},
  \bibinfo {author} {\bibfnamefont {Y.}~\bibnamefont {Saisyu}}, \bibinfo
  {author} {\bibfnamefont {H.}~\bibnamefont {Miyazaki}}, \bibinfo {author}
  {\bibfnamefont {S.}~\bibnamefont {Kimura}}, \bibinfo {author} {\bibfnamefont
  {T.}~\bibnamefont {Okuda}}, \bibinfo {author} {\bibfnamefont
  {I.}~\bibnamefont {Matsuda}}, \bibinfo {author} {\bibfnamefont
  {S.}~\bibnamefont {Murakami}},\ and\ \bibinfo {author} {\bibfnamefont
  {S.}~\bibnamefont {Hasegawa}},\ }\bibfield  {title} {\bibinfo {title}
  {Topological metal at the surface of an ultrathin
  ${\text{bi}}_{1\ensuremath{-}x}{\text{sb}}_{x}$ alloy film},\ }\href
  {https://doi.org/10.1103/PhysRevB.81.165422} {\bibfield  {journal} {\bibinfo
  {journal} {Phys. Rev. B}\ }\textbf {\bibinfo {volume} {81}},\ \bibinfo
  {pages} {165422} (\bibinfo {year} {2010})}\BibitemShut {NoStop}%
\bibitem [{\citenamefont {Nishide}\ \emph {et~al.}(2010)\citenamefont
  {Nishide}, \citenamefont {Taskin}, \citenamefont {Takeichi}, \citenamefont
  {Okuda}, \citenamefont {Kakizaki}, \citenamefont {Hirahara}, \citenamefont
  {Nakatsuji}, \citenamefont {Komori}, \citenamefont {Ando},\ and\
  \citenamefont {Matsuda}}]{Nishide2010}%
  \BibitemOpen
  \bibfield  {author} {\bibinfo {author} {\bibfnamefont {A.}~\bibnamefont
  {Nishide}}, \bibinfo {author} {\bibfnamefont {A.~A.}\ \bibnamefont {Taskin}},
  \bibinfo {author} {\bibfnamefont {Y.}~\bibnamefont {Takeichi}}, \bibinfo
  {author} {\bibfnamefont {T.}~\bibnamefont {Okuda}}, \bibinfo {author}
  {\bibfnamefont {A.}~\bibnamefont {Kakizaki}}, \bibinfo {author}
  {\bibfnamefont {T.}~\bibnamefont {Hirahara}}, \bibinfo {author}
  {\bibfnamefont {K.}~\bibnamefont {Nakatsuji}}, \bibinfo {author}
  {\bibfnamefont {F.}~\bibnamefont {Komori}}, \bibinfo {author} {\bibfnamefont
  {Y.}~\bibnamefont {Ando}},\ and\ \bibinfo {author} {\bibfnamefont
  {I.}~\bibnamefont {Matsuda}},\ }\bibfield  {title} {\bibinfo {title} {Direct
  mapping of the spin-filtered surface bands of a three-dimensional quantum
  spin hall insulator},\ }\href {https://doi.org/10.1103/PhysRevB.81.041309}
  {\bibfield  {journal} {\bibinfo  {journal} {Phys. Rev. B}\ }\textbf {\bibinfo
  {volume} {81}},\ \bibinfo {pages} {041309} (\bibinfo {year}
  {2010})}\BibitemShut {NoStop}%
\bibitem [{\citenamefont {Guo}\ \emph {et~al.}(2011)\citenamefont {Guo},
  \citenamefont {Sugawara}, \citenamefont {Takayama}, \citenamefont {Souma},
  \citenamefont {Sato}, \citenamefont {Satoh}, \citenamefont {Ohnishi},
  \citenamefont {Kitaura}, \citenamefont {Sasaki}, \citenamefont {Xue},\ and\
  \citenamefont {Takahashi}}]{HGuo2011}%
  \BibitemOpen
  \bibfield  {author} {\bibinfo {author} {\bibfnamefont {H.}~\bibnamefont
  {Guo}}, \bibinfo {author} {\bibfnamefont {K.}~\bibnamefont {Sugawara}},
  \bibinfo {author} {\bibfnamefont {A.}~\bibnamefont {Takayama}}, \bibinfo
  {author} {\bibfnamefont {S.}~\bibnamefont {Souma}}, \bibinfo {author}
  {\bibfnamefont {T.}~\bibnamefont {Sato}}, \bibinfo {author} {\bibfnamefont
  {N.}~\bibnamefont {Satoh}}, \bibinfo {author} {\bibfnamefont
  {A.}~\bibnamefont {Ohnishi}}, \bibinfo {author} {\bibfnamefont
  {M.}~\bibnamefont {Kitaura}}, \bibinfo {author} {\bibfnamefont
  {M.}~\bibnamefont {Sasaki}}, \bibinfo {author} {\bibfnamefont {Q.-K.}\
  \bibnamefont {Xue}},\ and\ \bibinfo {author} {\bibfnamefont {T.}~\bibnamefont
  {Takahashi}},\ }\bibfield  {title} {\bibinfo {title} {Evolution of surface
  states in bi${}_{1-x}$sb${}_{x}$ alloys across the topological phase
  transition},\ }\href {https://doi.org/10.1103/PhysRevB.83.201104} {\bibfield
  {journal} {\bibinfo  {journal} {Phys. Rev. B}\ }\textbf {\bibinfo {volume}
  {83}},\ \bibinfo {pages} {201104} (\bibinfo {year} {2011})}\BibitemShut
  {NoStop}%
\bibitem [{\citenamefont {Nakamura}\ \emph {et~al.}(2011)\citenamefont
  {Nakamura}, \citenamefont {Kousa}, \citenamefont {Taskin}, \citenamefont
  {Takeichi}, \citenamefont {Nishide}, \citenamefont {Kakizaki}, \citenamefont
  {D'Angelo}, \citenamefont {Lefevre}, \citenamefont {Bertran}, \citenamefont
  {Taleb-Ibrahimi}, \citenamefont {Komori}, \citenamefont {Kimura},
  \citenamefont {Kondo}, \citenamefont {Ando},\ and\ \citenamefont
  {Matsuda}}]{Nakamura2011}%
  \BibitemOpen
  \bibfield  {author} {\bibinfo {author} {\bibfnamefont {F.}~\bibnamefont
  {Nakamura}}, \bibinfo {author} {\bibfnamefont {Y.}~\bibnamefont {Kousa}},
  \bibinfo {author} {\bibfnamefont {A.~A.}\ \bibnamefont {Taskin}}, \bibinfo
  {author} {\bibfnamefont {Y.}~\bibnamefont {Takeichi}}, \bibinfo {author}
  {\bibfnamefont {A.}~\bibnamefont {Nishide}}, \bibinfo {author} {\bibfnamefont
  {A.}~\bibnamefont {Kakizaki}}, \bibinfo {author} {\bibfnamefont
  {M.}~\bibnamefont {D'Angelo}}, \bibinfo {author} {\bibfnamefont
  {P.}~\bibnamefont {Lefevre}}, \bibinfo {author} {\bibfnamefont
  {F.}~\bibnamefont {Bertran}}, \bibinfo {author} {\bibfnamefont
  {A.}~\bibnamefont {Taleb-Ibrahimi}}, \bibinfo {author} {\bibfnamefont
  {F.}~\bibnamefont {Komori}}, \bibinfo {author} {\bibfnamefont {S.-i.}\
  \bibnamefont {Kimura}}, \bibinfo {author} {\bibfnamefont {H.}~\bibnamefont
  {Kondo}}, \bibinfo {author} {\bibfnamefont {Y.}~\bibnamefont {Ando}},\ and\
  \bibinfo {author} {\bibfnamefont {I.}~\bibnamefont {Matsuda}},\ }\bibfield
  {title} {\bibinfo {title} {Topological transition in bi$_{1-x}$sb$_{x}$
  studied as a function of sb doping},\ }\href
  {https://doi.org/10.1103/PhysRevB.84.235308} {\bibfield  {journal} {\bibinfo
  {journal} {Phys. Rev. B}\ }\textbf {\bibinfo {volume} {84}},\ \bibinfo
  {pages} {235308} (\bibinfo {year} {2011})}\BibitemShut {NoStop}%
\bibitem [{\citenamefont {Ohtsubo}\ \emph {et~al.}(2013)\citenamefont
  {Ohtsubo}, \citenamefont {Perfetti}, \citenamefont {Goerbig}, \citenamefont
  {Fevre}, \citenamefont {Bertran},\ and\ \citenamefont
  {Taleb-Ibrahimi}}]{Ohtsubo2013}%
  \BibitemOpen
  \bibfield  {author} {\bibinfo {author} {\bibfnamefont {Y.}~\bibnamefont
  {Ohtsubo}}, \bibinfo {author} {\bibfnamefont {L.}~\bibnamefont {Perfetti}},
  \bibinfo {author} {\bibfnamefont {M.~O.}\ \bibnamefont {Goerbig}}, \bibinfo
  {author} {\bibfnamefont {P.~L.}\ \bibnamefont {Fevre}}, \bibinfo {author}
  {\bibfnamefont {F.}~\bibnamefont {Bertran}},\ and\ \bibinfo {author}
  {\bibfnamefont {A.}~\bibnamefont {Taleb-Ibrahimi}},\ }\bibfield  {title}
  {\bibinfo {title} {Non-trivial surface-band dispersion on bi(111)},\ }\href
  {http://stacks.iop.org/1367-2630/15/i=3/a=033041} {\bibfield  {journal}
  {\bibinfo  {journal} {New Journal of Physics}\ }\textbf {\bibinfo {volume}
  {15}},\ \bibinfo {pages} {033041} (\bibinfo {year} {2013})}\BibitemShut
  {NoStop}%
\bibitem [{\citenamefont {Benia}\ \emph {et~al.}(2015)\citenamefont {Benia},
  \citenamefont {Stra\ss{}er}, \citenamefont {Kern},\ and\ \citenamefont
  {Ast}}]{Benia2015}%
  \BibitemOpen
  \bibfield  {author} {\bibinfo {author} {\bibfnamefont {H.~M.}\ \bibnamefont
  {Benia}}, \bibinfo {author} {\bibfnamefont {C.}~\bibnamefont {Stra\ss{}er}},
  \bibinfo {author} {\bibfnamefont {K.}~\bibnamefont {Kern}},\ and\ \bibinfo
  {author} {\bibfnamefont {C.~R.}\ \bibnamefont {Ast}},\ }\bibfield  {title}
  {\bibinfo {title} {Surface band structure of
  ${\mathrm{bi}}_{1\ensuremath{-}x}{\mathrm{sb}}_{x}(111)$},\ }\href
  {https://doi.org/10.1103/PhysRevB.91.161406} {\bibfield  {journal} {\bibinfo
  {journal} {Phys. Rev. B}\ }\textbf {\bibinfo {volume} {91}},\ \bibinfo
  {pages} {161406} (\bibinfo {year} {2015})}\BibitemShut {NoStop}%
\bibitem [{\citenamefont {Ito}\ \emph {et~al.}(2016)\citenamefont {Ito},
  \citenamefont {Feng}, \citenamefont {Arita}, \citenamefont {Takayama},
  \citenamefont {Liu}, \citenamefont {Someya}, \citenamefont {Chen},
  \citenamefont {Iimori}, \citenamefont {Namatame}, \citenamefont {Taniguchi},
  \citenamefont {Cheng}, \citenamefont {Tang}, \citenamefont {Komori},
  \citenamefont {Kobayashi}, \citenamefont {Chiang},\ and\ \citenamefont
  {Matsuda}}]{Ito2016}%
  \BibitemOpen
  \bibfield  {author} {\bibinfo {author} {\bibfnamefont {S.}~\bibnamefont
  {Ito}}, \bibinfo {author} {\bibfnamefont {B.}~\bibnamefont {Feng}}, \bibinfo
  {author} {\bibfnamefont {M.}~\bibnamefont {Arita}}, \bibinfo {author}
  {\bibfnamefont {A.}~\bibnamefont {Takayama}}, \bibinfo {author}
  {\bibfnamefont {R.-Y.}\ \bibnamefont {Liu}}, \bibinfo {author} {\bibfnamefont
  {T.}~\bibnamefont {Someya}}, \bibinfo {author} {\bibfnamefont {W.-C.}\
  \bibnamefont {Chen}}, \bibinfo {author} {\bibfnamefont {T.}~\bibnamefont
  {Iimori}}, \bibinfo {author} {\bibfnamefont {H.}~\bibnamefont {Namatame}},
  \bibinfo {author} {\bibfnamefont {M.}~\bibnamefont {Taniguchi}}, \bibinfo
  {author} {\bibfnamefont {C.-M.}\ \bibnamefont {Cheng}}, \bibinfo {author}
  {\bibfnamefont {S.-J.}\ \bibnamefont {Tang}}, \bibinfo {author}
  {\bibfnamefont {F.}~\bibnamefont {Komori}}, \bibinfo {author} {\bibfnamefont
  {K.}~\bibnamefont {Kobayashi}}, \bibinfo {author} {\bibfnamefont {T.-C.}\
  \bibnamefont {Chiang}},\ and\ \bibinfo {author} {\bibfnamefont
  {I.}~\bibnamefont {Matsuda}},\ }\bibfield  {title} {\bibinfo {title} {Proving
  nontrivial topology of pure bismuth by quantum confinement},\ }\href
  {https://doi.org/10.1103/PhysRevLett.117.236402} {\bibfield  {journal}
  {\bibinfo  {journal} {Phys. Rev. Lett.}\ }\textbf {\bibinfo {volume} {117}},\
  \bibinfo {pages} {236402} (\bibinfo {year} {2016})}\BibitemShut {NoStop}%
\bibitem [{\citenamefont {Ohtsubo}\ and\ \citenamefont
  {Kimura}(2016)}]{Ohtsubo2016}%
  \BibitemOpen
  \bibfield  {author} {\bibinfo {author} {\bibfnamefont {Y.}~\bibnamefont
  {Ohtsubo}}\ and\ \bibinfo {author} {\bibfnamefont {S.}~\bibnamefont
  {Kimura}},\ }\bibfield  {title} {\bibinfo {title} {Topological phase
  transition of single-crystal bi based on empirical tight-binding
  calculations},\ }\href {http://stacks.iop.org/1367-2630/18/i=12/a=123015}
  {\bibfield  {journal} {\bibinfo  {journal} {New Journal of Physics}\ }\textbf
  {\bibinfo {volume} {18}},\ \bibinfo {pages} {123015} (\bibinfo {year}
  {2016})}\BibitemShut {NoStop}%
\bibitem [{\citenamefont {Fukushima}\ \emph {et~al.}(2023)\citenamefont
  {Fukushima}, \citenamefont {Kawaguchi}, \citenamefont {Kuroda}, \citenamefont
  {Ochi}, \citenamefont {Tanaka}, \citenamefont {Harasawa}, \citenamefont
  {Iimori}, \citenamefont {Zhao}, \citenamefont {Tani}, \citenamefont {Yaji},
  \citenamefont {Shin}, \citenamefont {Komori}, \citenamefont {Kobayashi},\
  and\ \citenamefont {Kondo}}]{Fukushima2023}%
  \BibitemOpen
  \bibfield  {author} {\bibinfo {author} {\bibfnamefont {Y.}~\bibnamefont
  {Fukushima}}, \bibinfo {author} {\bibfnamefont {K.}~\bibnamefont
  {Kawaguchi}}, \bibinfo {author} {\bibfnamefont {K.}~\bibnamefont {Kuroda}},
  \bibinfo {author} {\bibfnamefont {M.}~\bibnamefont {Ochi}}, \bibinfo {author}
  {\bibfnamefont {H.}~\bibnamefont {Tanaka}}, \bibinfo {author} {\bibfnamefont
  {A.}~\bibnamefont {Harasawa}}, \bibinfo {author} {\bibfnamefont
  {T.}~\bibnamefont {Iimori}}, \bibinfo {author} {\bibfnamefont
  {Z.}~\bibnamefont {Zhao}}, \bibinfo {author} {\bibfnamefont {S.}~\bibnamefont
  {Tani}}, \bibinfo {author} {\bibfnamefont {K.}~\bibnamefont {Yaji}}, \bibinfo
  {author} {\bibfnamefont {S.}~\bibnamefont {Shin}}, \bibinfo {author}
  {\bibfnamefont {F.}~\bibnamefont {Komori}}, \bibinfo {author} {\bibfnamefont
  {Y.}~\bibnamefont {Kobayashi}},\ and\ \bibinfo {author} {\bibfnamefont
  {T.}~\bibnamefont {Kondo}},\ }\href {https://arxiv.org/abs/2303.17816}
  {\bibinfo {title} {Spin-polarized saddle points in the topological surface
  states of the elemental bismuth revealed by a pump-probe spin-resolved
  arpes}} (\bibinfo {year} {2023}),\ \Eprint {https://arxiv.org/abs/2303.17816}
  {arXiv:2303.17816 [cond-mat.mtrl-sci]} \BibitemShut {NoStop}%
\bibitem [{\citenamefont {Teo}\ \emph {et~al.}(2008)\citenamefont {Teo},
  \citenamefont {Fu},\ and\ \citenamefont {Kane}}]{Teo2008}%
  \BibitemOpen
  \bibfield  {author} {\bibinfo {author} {\bibfnamefont {J.~C.~Y.}\
  \bibnamefont {Teo}}, \bibinfo {author} {\bibfnamefont {L.}~\bibnamefont
  {Fu}},\ and\ \bibinfo {author} {\bibfnamefont {C.~L.}\ \bibnamefont {Kane}},\
  }\bibfield  {title} {\bibinfo {title} {Surface states and topological
  invariants in three-dimensional topological insulators: Application to
  bi$_{1-x}$sb$_x$},\ }\href {https://doi.org/10.1103/PhysRevB.78.045426}
  {\bibfield  {journal} {\bibinfo  {journal} {Phys. Rev. B}\ }\textbf {\bibinfo
  {volume} {78}},\ \bibinfo {pages} {045426} (\bibinfo {year}
  {2008})}\BibitemShut {NoStop}%
\bibitem [{\citenamefont {Zhang}\ \emph {et~al.}(2009)\citenamefont {Zhang},
  \citenamefont {Liu}, \citenamefont {Qi}, \citenamefont {Deng}, \citenamefont
  {Dai}, \citenamefont {Zhang},\ and\ \citenamefont {Fang}}]{HJZhang2009}%
  \BibitemOpen
  \bibfield  {author} {\bibinfo {author} {\bibfnamefont {H.-J.}\ \bibnamefont
  {Zhang}}, \bibinfo {author} {\bibfnamefont {C.-X.}\ \bibnamefont {Liu}},
  \bibinfo {author} {\bibfnamefont {X.-L.}\ \bibnamefont {Qi}}, \bibinfo
  {author} {\bibfnamefont {X.-Y.}\ \bibnamefont {Deng}}, \bibinfo {author}
  {\bibfnamefont {X.}~\bibnamefont {Dai}}, \bibinfo {author} {\bibfnamefont
  {S.-C.}\ \bibnamefont {Zhang}},\ and\ \bibinfo {author} {\bibfnamefont
  {Z.}~\bibnamefont {Fang}},\ }\bibfield  {title} {\bibinfo {title} {Electronic
  structures and surface states of the topological insulator
  ${\text{bi}}_{1\ensuremath{-}x}{\text{sb}}_{x}$},\ }\href
  {https://doi.org/10.1103/PhysRevB.80.085307} {\bibfield  {journal} {\bibinfo
  {journal} {Phys. Rev. B}\ }\textbf {\bibinfo {volume} {80}},\ \bibinfo
  {pages} {085307} (\bibinfo {year} {2009})}\BibitemShut {NoStop}%
\bibitem [{\citenamefont {Aguilera}\ \emph {et~al.}(2015)\citenamefont
  {Aguilera}, \citenamefont {Friedrich},\ and\ \citenamefont
  {Bl\"ugel}}]{Aguilera2015}%
  \BibitemOpen
  \bibfield  {author} {\bibinfo {author} {\bibfnamefont {I.}~\bibnamefont
  {Aguilera}}, \bibinfo {author} {\bibfnamefont {C.}~\bibnamefont
  {Friedrich}},\ and\ \bibinfo {author} {\bibfnamefont {S.}~\bibnamefont
  {Bl\"ugel}},\ }\bibfield  {title} {\bibinfo {title} {Electronic phase
  transitions of bismuth under strain from relativistic self-consistent $gw$
  calculations},\ }\href {https://doi.org/10.1103/PhysRevB.91.125129}
  {\bibfield  {journal} {\bibinfo  {journal} {Phys. Rev. B}\ }\textbf {\bibinfo
  {volume} {91}},\ \bibinfo {pages} {125129} (\bibinfo {year}
  {2015})}\BibitemShut {NoStop}%
\bibitem [{\citenamefont {Fuseya}\ and\ \citenamefont
  {Fukuyama}(2018)}]{Fuseya2018}%
  \BibitemOpen
  \bibfield  {author} {\bibinfo {author} {\bibfnamefont {Y.}~\bibnamefont
  {Fuseya}}\ and\ \bibinfo {author} {\bibfnamefont {H.}~\bibnamefont
  {Fukuyama}},\ }\bibfield  {title} {\bibinfo {title} {Analytical solutions for
  the surface states of bi$_{1-x}$sb$_x$},\ }\href
  {https://doi.org/10.7566/JPSJ.87.044710} {\bibfield  {journal} {\bibinfo
  {journal} {J. Phys. Soc. Jpn.}\ }\textbf {\bibinfo {volume} {87}},\ \bibinfo
  {pages} {044710} (\bibinfo {year} {2018})}\BibitemShut {NoStop}%
\bibitem [{\citenamefont {Aguilera}\ \emph {et~al.}(2021)\citenamefont
  {Aguilera}, \citenamefont {Kim}, \citenamefont {Friedrich}, \citenamefont
  {Bihlmayer},\ and\ \citenamefont {Bl\"ugel}}]{Aguilera2021}%
  \BibitemOpen
  \bibfield  {author} {\bibinfo {author} {\bibfnamefont {I.}~\bibnamefont
  {Aguilera}}, \bibinfo {author} {\bibfnamefont {H.-J.}\ \bibnamefont {Kim}},
  \bibinfo {author} {\bibfnamefont {C.}~\bibnamefont {Friedrich}}, \bibinfo
  {author} {\bibfnamefont {G.}~\bibnamefont {Bihlmayer}},\ and\ \bibinfo
  {author} {\bibfnamefont {S.}~\bibnamefont {Bl\"ugel}},\ }\bibfield  {title}
  {\bibinfo {title} {$z_2$ topology of bismuth},\ }\href
  {https://doi.org/10.1103/PhysRevMaterials.5.L091201} {\bibfield  {journal}
  {\bibinfo  {journal} {Phys. Rev. Materials}\ }\textbf {\bibinfo {volume}
  {5}},\ \bibinfo {pages} {L091201} (\bibinfo {year} {2021})}\BibitemShut
  {NoStop}%
\bibitem [{\citenamefont {Ferreira}(1967)}]{Ferreira1967}%
  \BibitemOpen
  \bibfield  {author} {\bibinfo {author} {\bibfnamefont {L.~G.}\ \bibnamefont
  {Ferreira}},\ }\bibfield  {title} {\bibinfo {title} {Relativistic band
  structure calculation for bismuth},\ }\href
  {https://doi.org/http://dx.doi.org/10.1016/0022-3697(67)90166-7} {\bibfield
  {journal} {\bibinfo  {journal} {Journal of Physics and Chemistry of Solids}\
  }\textbf {\bibinfo {volume} {28}},\ \bibinfo {pages} {1891 } (\bibinfo {year}
  {1967})}\BibitemShut {NoStop}%
\bibitem [{\citenamefont {Ferreira}(1968)}]{Ferreira1968}%
  \BibitemOpen
  \bibfield  {author} {\bibinfo {author} {\bibfnamefont {L.~G.}\ \bibnamefont
  {Ferreira}},\ }\bibfield  {title} {\bibinfo {title} {Band structure
  calculation for bismuth: Comparison with experiment},\ }\href
  {https://doi.org/http://dx.doi.org/10.1016/0022-3697(68)90081-4} {\bibfield
  {journal} {\bibinfo  {journal} {Journal of Physics and Chemistry of Solids}\
  }\textbf {\bibinfo {volume} {29}},\ \bibinfo {pages} {357 } (\bibinfo {year}
  {1968})}\BibitemShut {NoStop}%
\bibitem [{\citenamefont {Golin}(1968)}]{Golin1968}%
  \BibitemOpen
  \bibfield  {author} {\bibinfo {author} {\bibfnamefont {S.}~\bibnamefont
  {Golin}},\ }\bibfield  {title} {\bibinfo {title} {Band structure of bismuth:
  Pseudopotential approach},\ }\href {https://doi.org/10.1103/PhysRev.166.643}
  {\bibfield  {journal} {\bibinfo  {journal} {Phys. Rev.}\ }\textbf {\bibinfo
  {volume} {166}},\ \bibinfo {pages} {643} (\bibinfo {year}
  {1968})}\BibitemShut {NoStop}%
\bibitem [{\citenamefont {Gonze}\ \emph {et~al.}(1988)\citenamefont {Gonze},
  \citenamefont {Michenaud},\ and\ \citenamefont {Vigneron}}]{Gonze1988}%
  \BibitemOpen
  \bibfield  {author} {\bibinfo {author} {\bibfnamefont {X.}~\bibnamefont
  {Gonze}}, \bibinfo {author} {\bibfnamefont {J.-P.}\ \bibnamefont
  {Michenaud}},\ and\ \bibinfo {author} {\bibfnamefont {J.-P.}\ \bibnamefont
  {Vigneron}},\ }\bibfield  {title} {\bibinfo {title} {Ab initio calculations
  of bismuth properties, including spin-orbit coupling},\ }\href@noop {}
  {\bibfield  {journal} {\bibinfo  {journal} {Physica Scripta}\ }\textbf
  {\bibinfo {volume} {37}},\ \bibinfo {pages} {785} (\bibinfo {year}
  {1988})}\BibitemShut {NoStop}%
\bibitem [{\citenamefont {Gonze}\ \emph {et~al.}(1990)\citenamefont {Gonze},
  \citenamefont {Michenaud},\ and\ \citenamefont {Vigneron}}]{Gonze1990}%
  \BibitemOpen
  \bibfield  {author} {\bibinfo {author} {\bibfnamefont {X.}~\bibnamefont
  {Gonze}}, \bibinfo {author} {\bibfnamefont {J.-P.}\ \bibnamefont
  {Michenaud}},\ and\ \bibinfo {author} {\bibfnamefont {J.-P.}\ \bibnamefont
  {Vigneron}},\ }\bibfield  {title} {\bibinfo {title} {First-principles study
  of as, sb, and bi electronic properties},\ }\href
  {https://doi.org/10.1103/PhysRevB.41.11827} {\bibfield  {journal} {\bibinfo
  {journal} {Phys. Rev. B}\ }\textbf {\bibinfo {volume} {41}},\ \bibinfo
  {pages} {11827} (\bibinfo {year} {1990})}\BibitemShut {NoStop}%
\bibitem [{\citenamefont {Shick}\ \emph {et~al.}(1999)\citenamefont {Shick},
  \citenamefont {Ketterson}, \citenamefont {Novikov},\ and\ \citenamefont
  {Freeman}}]{Shick1999}%
  \BibitemOpen
  \bibfield  {author} {\bibinfo {author} {\bibfnamefont {A.~B.}\ \bibnamefont
  {Shick}}, \bibinfo {author} {\bibfnamefont {J.~B.}\ \bibnamefont
  {Ketterson}}, \bibinfo {author} {\bibfnamefont {D.~L.}\ \bibnamefont
  {Novikov}},\ and\ \bibinfo {author} {\bibfnamefont {A.~J.}\ \bibnamefont
  {Freeman}},\ }\bibfield  {title} {\bibinfo {title} {Electronic structure,
  phase stability, and semimetal-semiconductor transitions in bi},\ }\href
  {https://doi.org/10.1103/PhysRevB.60.15484} {\bibfield  {journal} {\bibinfo
  {journal} {Phys. Rev. B}\ }\textbf {\bibinfo {volume} {60}},\ \bibinfo
  {pages} {15484} (\bibinfo {year} {1999})}\BibitemShut {NoStop}%
\bibitem [{\citenamefont {Timrov}\ \emph {et~al.}(2012)\citenamefont {Timrov},
  \citenamefont {Kampfrath}, \citenamefont {Faure}, \citenamefont {Vast},
  \citenamefont {Ast}, \citenamefont {Frischkorn}, \citenamefont {Wolf},
  \citenamefont {Gava},\ and\ \citenamefont {Perfetti}}]{Timrov2012}%
  \BibitemOpen
  \bibfield  {author} {\bibinfo {author} {\bibfnamefont {I.}~\bibnamefont
  {Timrov}}, \bibinfo {author} {\bibfnamefont {T.}~\bibnamefont {Kampfrath}},
  \bibinfo {author} {\bibfnamefont {J.}~\bibnamefont {Faure}}, \bibinfo
  {author} {\bibfnamefont {N.}~\bibnamefont {Vast}}, \bibinfo {author}
  {\bibfnamefont {C.~R.}\ \bibnamefont {Ast}}, \bibinfo {author} {\bibfnamefont
  {C.}~\bibnamefont {Frischkorn}}, \bibinfo {author} {\bibfnamefont
  {M.}~\bibnamefont {Wolf}}, \bibinfo {author} {\bibfnamefont {P.}~\bibnamefont
  {Gava}},\ and\ \bibinfo {author} {\bibfnamefont {L.}~\bibnamefont
  {Perfetti}},\ }\bibfield  {title} {\bibinfo {title} {Thermalization of
  photoexcited carriers in bismuth investigated by time-resolved terahertz
  spectroscopy and ab initio calculations},\ }\href
  {https://doi.org/10.1103/PhysRevB.85.155139} {\bibfield  {journal} {\bibinfo
  {journal} {Phys. Rev. B}\ }\textbf {\bibinfo {volume} {85}},\ \bibinfo
  {pages} {155139} (\bibinfo {year} {2012})}\BibitemShut {NoStop}%
\bibitem [{\citenamefont {Zhou}\ \emph {et~al.}(2008)\citenamefont {Zhou},
  \citenamefont {Lu}, \citenamefont {Chu}, \citenamefont {Shen},\ and\
  \citenamefont {Niu}}]{BZhou2008}%
  \BibitemOpen
  \bibfield  {author} {\bibinfo {author} {\bibfnamefont {B.}~\bibnamefont
  {Zhou}}, \bibinfo {author} {\bibfnamefont {H.-Z.}\ \bibnamefont {Lu}},
  \bibinfo {author} {\bibfnamefont {R.-L.}\ \bibnamefont {Chu}}, \bibinfo
  {author} {\bibfnamefont {S.-Q.}\ \bibnamefont {Shen}},\ and\ \bibinfo
  {author} {\bibfnamefont {Q.}~\bibnamefont {Niu}},\ }\bibfield  {title}
  {\bibinfo {title} {Finite size effects on helical edge states in a quantum
  spin-hall system},\ }\href {https://doi.org/10.1103/PhysRevLett.101.246807}
  {\bibfield  {journal} {\bibinfo  {journal} {Phys. Rev. Lett.}\ }\textbf
  {\bibinfo {volume} {101}},\ \bibinfo {pages} {246807} (\bibinfo {year}
  {2008})}\BibitemShut {NoStop}%
\bibitem [{\citenamefont {Linder}\ \emph {et~al.}(2009)\citenamefont {Linder},
  \citenamefont {Yokoyama},\ and\ \citenamefont {Sudb\o{}}}]{Linder2009}%
  \BibitemOpen
  \bibfield  {author} {\bibinfo {author} {\bibfnamefont {J.}~\bibnamefont
  {Linder}}, \bibinfo {author} {\bibfnamefont {T.}~\bibnamefont {Yokoyama}},\
  and\ \bibinfo {author} {\bibfnamefont {A.}~\bibnamefont {Sudb\o{}}},\
  }\bibfield  {title} {\bibinfo {title} {Anomalous finite size effects on
  surface states in the topological insulator
  ${\text{bi}}_{2}{\text{se}}_{3}$},\ }\href
  {https://doi.org/10.1103/PhysRevB.80.205401} {\bibfield  {journal} {\bibinfo
  {journal} {Phys. Rev. B}\ }\textbf {\bibinfo {volume} {80}},\ \bibinfo
  {pages} {205401} (\bibinfo {year} {2009})}\BibitemShut {NoStop}%
\bibitem [{\citenamefont {Lu}\ \emph {et~al.}(2010)\citenamefont {Lu},
  \citenamefont {Shan}, \citenamefont {Yao}, \citenamefont {Niu},\ and\
  \citenamefont {Shen}}]{HZLu2010}%
  \BibitemOpen
  \bibfield  {author} {\bibinfo {author} {\bibfnamefont {H.-Z.}\ \bibnamefont
  {Lu}}, \bibinfo {author} {\bibfnamefont {W.-Y.}\ \bibnamefont {Shan}},
  \bibinfo {author} {\bibfnamefont {W.}~\bibnamefont {Yao}}, \bibinfo {author}
  {\bibfnamefont {Q.}~\bibnamefont {Niu}},\ and\ \bibinfo {author}
  {\bibfnamefont {S.-Q.}\ \bibnamefont {Shen}},\ }\bibfield  {title} {\bibinfo
  {title} {Massive dirac fermions and spin physics in an ultrathin film of
  topological insulator},\ }\href {https://doi.org/10.1103/PhysRevB.81.115407}
  {\bibfield  {journal} {\bibinfo  {journal} {Phys. Rev. B}\ }\textbf {\bibinfo
  {volume} {81}},\ \bibinfo {pages} {115407} (\bibinfo {year}
  {2010})}\BibitemShut {NoStop}%
\bibitem [{\citenamefont {Shen}(2012)}]{SQShen_book}%
  \BibitemOpen
  \bibfield  {author} {\bibinfo {author} {\bibfnamefont {S.-Q.}\ \bibnamefont
  {Shen}},\ }\href@noop {} {\emph {\bibinfo {title} {Topological Insulators}}}\
  (\bibinfo  {publisher} {Springer-Verlag Berlin Heidelberg},\ \bibinfo {year}
  {2012})\BibitemShut {NoStop}%
\bibitem [{\citenamefont {Ozawa}\ \emph {et~al.}(2014)\citenamefont {Ozawa},
  \citenamefont {Yamakage}, \citenamefont {Sato},\ and\ \citenamefont
  {Tanaka}}]{Ozawa2014}%
  \BibitemOpen
  \bibfield  {author} {\bibinfo {author} {\bibfnamefont {H.}~\bibnamefont
  {Ozawa}}, \bibinfo {author} {\bibfnamefont {A.}~\bibnamefont {Yamakage}},
  \bibinfo {author} {\bibfnamefont {M.}~\bibnamefont {Sato}},\ and\ \bibinfo
  {author} {\bibfnamefont {Y.}~\bibnamefont {Tanaka}},\ }\bibfield  {title}
  {\bibinfo {title} {Topological phase transition in a topological crystalline
  insulator induced by finite-size effects},\ }\href
  {https://doi.org/10.1103/PhysRevB.90.045309} {\bibfield  {journal} {\bibinfo
  {journal} {Phys. Rev. B}\ }\textbf {\bibinfo {volume} {90}},\ \bibinfo
  {pages} {045309} (\bibinfo {year} {2014})}\BibitemShut {NoStop}%
\bibitem [{\citenamefont {Schindler}\ \emph {et~al.}(2018)\citenamefont
  {Schindler}, \citenamefont {Wang}, \citenamefont {Vergniory}, \citenamefont
  {Cook}, \citenamefont {Murani}, \citenamefont {Sengupta}, \citenamefont
  {Kasumov}, \citenamefont {Deblock}, \citenamefont {Jeon}, \citenamefont
  {Drozdov}, \citenamefont {Bouchiat}, \citenamefont {Gu{\'e}ron},
  \citenamefont {Yazdani}, \citenamefont {Bernevig},\ and\ \citenamefont
  {Neupert}}]{Schindler2018b}%
  \BibitemOpen
  \bibfield  {author} {\bibinfo {author} {\bibfnamefont {F.}~\bibnamefont
  {Schindler}}, \bibinfo {author} {\bibfnamefont {Z.}~\bibnamefont {Wang}},
  \bibinfo {author} {\bibfnamefont {M.~G.}\ \bibnamefont {Vergniory}}, \bibinfo
  {author} {\bibfnamefont {A.~M.}\ \bibnamefont {Cook}}, \bibinfo {author}
  {\bibfnamefont {A.}~\bibnamefont {Murani}}, \bibinfo {author} {\bibfnamefont
  {S.}~\bibnamefont {Sengupta}}, \bibinfo {author} {\bibfnamefont {A.~Y.}\
  \bibnamefont {Kasumov}}, \bibinfo {author} {\bibfnamefont {R.}~\bibnamefont
  {Deblock}}, \bibinfo {author} {\bibfnamefont {S.}~\bibnamefont {Jeon}},
  \bibinfo {author} {\bibfnamefont {I.}~\bibnamefont {Drozdov}}, \bibinfo
  {author} {\bibfnamefont {H.}~\bibnamefont {Bouchiat}}, \bibinfo {author}
  {\bibfnamefont {S.}~\bibnamefont {Gu{\'e}ron}}, \bibinfo {author}
  {\bibfnamefont {A.}~\bibnamefont {Yazdani}}, \bibinfo {author} {\bibfnamefont
  {B.~A.}\ \bibnamefont {Bernevig}},\ and\ \bibinfo {author} {\bibfnamefont
  {T.}~\bibnamefont {Neupert}},\ }\bibfield  {title} {\bibinfo {title}
  {Higher-order topology in bismuth},\ }\href
  {https://doi.org/10.1038/s41567-018-0224-7} {\bibfield  {journal} {\bibinfo
  {journal} {Nature Physics}\ }\textbf {\bibinfo {volume} {14}},\ \bibinfo
  {pages} {918} (\bibinfo {year} {2018})}\BibitemShut {NoStop}%
\bibitem [{\citenamefont {Aggarwal}\ \emph {et~al.}(2021)\citenamefont
  {Aggarwal}, \citenamefont {Zhu}, \citenamefont {Hughes},\ and\ \citenamefont
  {Madhavan}}]{Aggarwal2021}%
  \BibitemOpen
  \bibfield  {author} {\bibinfo {author} {\bibfnamefont {L.}~\bibnamefont
  {Aggarwal}}, \bibinfo {author} {\bibfnamefont {P.}~\bibnamefont {Zhu}},
  \bibinfo {author} {\bibfnamefont {T.~L.}\ \bibnamefont {Hughes}},\ and\
  \bibinfo {author} {\bibfnamefont {V.}~\bibnamefont {Madhavan}},\ }\bibfield
  {title} {\bibinfo {title} {Evidence for higher order topology in bi and
  bi0.92sb0.08},\ }\href {https://doi.org/10.1038/s41467-021-24683-8}
  {\bibfield  {journal} {\bibinfo  {journal} {Nature Communications}\ }\textbf
  {\bibinfo {volume} {12}},\ \bibinfo {pages} {4420} (\bibinfo {year}
  {2021})}\BibitemShut {NoStop}%
\bibitem [{\citenamefont {Hirahara}\ \emph {et~al.}(2012)\citenamefont
  {Hirahara}, \citenamefont {Fukui}, \citenamefont {Shirasawa}, \citenamefont
  {Yamada}, \citenamefont {Aitani}, \citenamefont {Miyazaki}, \citenamefont
  {Matsunami}, \citenamefont {Kimura}, \citenamefont {Takahashi}, \citenamefont
  {Hasegawa},\ and\ \citenamefont {Kobayashi}}]{Hirahara2012}%
  \BibitemOpen
  \bibfield  {author} {\bibinfo {author} {\bibfnamefont {T.}~\bibnamefont
  {Hirahara}}, \bibinfo {author} {\bibfnamefont {N.}~\bibnamefont {Fukui}},
  \bibinfo {author} {\bibfnamefont {T.}~\bibnamefont {Shirasawa}}, \bibinfo
  {author} {\bibfnamefont {M.}~\bibnamefont {Yamada}}, \bibinfo {author}
  {\bibfnamefont {M.}~\bibnamefont {Aitani}}, \bibinfo {author} {\bibfnamefont
  {H.}~\bibnamefont {Miyazaki}}, \bibinfo {author} {\bibfnamefont
  {M.}~\bibnamefont {Matsunami}}, \bibinfo {author} {\bibfnamefont
  {S.}~\bibnamefont {Kimura}}, \bibinfo {author} {\bibfnamefont
  {T.}~\bibnamefont {Takahashi}}, \bibinfo {author} {\bibfnamefont
  {S.}~\bibnamefont {Hasegawa}},\ and\ \bibinfo {author} {\bibfnamefont
  {K.}~\bibnamefont {Kobayashi}},\ }\bibfield  {title} {\bibinfo {title}
  {Atomic and electronic structure of ultrathin bi(111) films grown on
  ${\mathrm{bi}}_{2}{\mathrm{te}}_{3}(111)$ substrates: Evidence for a
  strain-induced topological phase transition},\ }\href
  {https://doi.org/10.1103/PhysRevLett.109.227401} {\bibfield  {journal}
  {\bibinfo  {journal} {Phys. Rev. Lett.}\ }\textbf {\bibinfo {volume} {109}},\
  \bibinfo {pages} {227401} (\bibinfo {year} {2012})}\BibitemShut {NoStop}%
\bibitem [{\citenamefont {Ohtsubo}\ \emph {et~al.}(2019)\citenamefont
  {Ohtsubo}, \citenamefont {Yamashita}, \citenamefont {Kishi}, \citenamefont
  {Ideta}, \citenamefont {Tanaka}, \citenamefont {Yamane}, \citenamefont
  {Rault}, \citenamefont {F{\`e}vre}, \citenamefont {Bertran},\ and\
  \citenamefont {Kimura}}]{Ohtsubo2019}%
  \BibitemOpen
  \bibfield  {author} {\bibinfo {author} {\bibfnamefont {Y.}~\bibnamefont
  {Ohtsubo}}, \bibinfo {author} {\bibfnamefont {Y.}~\bibnamefont {Yamashita}},
  \bibinfo {author} {\bibfnamefont {J.}~\bibnamefont {Kishi}}, \bibinfo
  {author} {\bibfnamefont {S.}~\bibnamefont {Ideta}}, \bibinfo {author}
  {\bibfnamefont {K.}~\bibnamefont {Tanaka}}, \bibinfo {author} {\bibfnamefont
  {H.}~\bibnamefont {Yamane}}, \bibinfo {author} {\bibfnamefont {J.~E.}\
  \bibnamefont {Rault}}, \bibinfo {author} {\bibfnamefont {P.~L.}\ \bibnamefont
  {F{\`e}vre}}, \bibinfo {author} {\bibfnamefont {F.}~\bibnamefont {Bertran}},\
  and\ \bibinfo {author} {\bibfnamefont {S.}~\bibnamefont {Kimura}},\
  }\bibfield  {title} {\bibinfo {title} {Temperature-driven modification of
  surface electronic structure on bismuth, a topological border material},\
  }\href {https://doi.org/10.1088/1361-6463/ab1515} {\bibfield  {journal}
  {\bibinfo  {journal} {Journal of Physics D: Applied Physics}\ }\textbf
  {\bibinfo {volume} {52}},\ \bibinfo {pages} {254002} (\bibinfo {year}
  {2019})}\BibitemShut {NoStop}%
\bibitem [{\citenamefont {M\"onig}\ \emph {et~al.}(2005)\citenamefont
  {M\"onig}, \citenamefont {Sun}, \citenamefont {Koroteev}, \citenamefont
  {Bihlmayer}, \citenamefont {Wells}, \citenamefont {Chulkov}, \citenamefont
  {Pohl},\ and\ \citenamefont {Hofmann}}]{Monig2005}%
  \BibitemOpen
  \bibfield  {author} {\bibinfo {author} {\bibfnamefont {H.}~\bibnamefont
  {M\"onig}}, \bibinfo {author} {\bibfnamefont {J.}~\bibnamefont {Sun}},
  \bibinfo {author} {\bibfnamefont {Y.~M.}\ \bibnamefont {Koroteev}}, \bibinfo
  {author} {\bibfnamefont {G.}~\bibnamefont {Bihlmayer}}, \bibinfo {author}
  {\bibfnamefont {J.}~\bibnamefont {Wells}}, \bibinfo {author} {\bibfnamefont
  {E.~V.}\ \bibnamefont {Chulkov}}, \bibinfo {author} {\bibfnamefont
  {K.}~\bibnamefont {Pohl}},\ and\ \bibinfo {author} {\bibfnamefont
  {P.}~\bibnamefont {Hofmann}},\ }\bibfield  {title} {\bibinfo {title}
  {Structure of the (111) surface of bismuth: Leed analysis and
  first-principles calculations},\ }\href
  {https://doi.org/10.1103/PhysRevB.72.085410} {\bibfield  {journal} {\bibinfo
  {journal} {Phys. Rev. B}\ }\textbf {\bibinfo {volume} {72}},\ \bibinfo
  {pages} {085410} (\bibinfo {year} {2005})}\BibitemShut {NoStop}%
\bibitem [{\citenamefont {Hofmann}(2006)}]{Hofmann2006}%
  \BibitemOpen
  \bibfield  {author} {\bibinfo {author} {\bibfnamefont {P.}~\bibnamefont
  {Hofmann}},\ }\bibfield  {title} {\bibinfo {title} {The surfaces of bismuth:
  Structural and electronic properties},\ }\href
  {https://doi.org/https://doi.org/10.1016/j.progsurf.2006.03.001} {\bibfield
  {journal} {\bibinfo  {journal} {Progress in Surface Science}\ }\textbf
  {\bibinfo {volume} {81}},\ \bibinfo {pages} {191} (\bibinfo {year}
  {2006})}\BibitemShut {NoStop}%
\bibitem [{\citenamefont {Saito}\ \emph {et~al.}(2016)\citenamefont {Saito},
  \citenamefont {Sawahata}, \citenamefont {Komine},\ and\ \citenamefont
  {Aono}}]{Saito2016}%
  \BibitemOpen
  \bibfield  {author} {\bibinfo {author} {\bibfnamefont {K.}~\bibnamefont
  {Saito}}, \bibinfo {author} {\bibfnamefont {H.}~\bibnamefont {Sawahata}},
  \bibinfo {author} {\bibfnamefont {T.}~\bibnamefont {Komine}},\ and\ \bibinfo
  {author} {\bibfnamefont {T.}~\bibnamefont {Aono}},\ }\bibfield  {title}
  {\bibinfo {title} {Tight-binding theory of surface spin states on bismuth
  thin films},\ }\href {https://doi.org/10.1103/PhysRevB.93.041301} {\bibfield
  {journal} {\bibinfo  {journal} {Phys. Rev. B}\ }\textbf {\bibinfo {volume}
  {93}},\ \bibinfo {pages} {041301} (\bibinfo {year} {2016})}\BibitemShut
  {NoStop}%
\bibitem [{\citenamefont {Asaka}\ \emph {et~al.}(2022)\citenamefont {Asaka},
  \citenamefont {Kikuchi},\ and\ \citenamefont {Fuseya}}]{Asaka2022}%
  \BibitemOpen
  \bibfield  {author} {\bibinfo {author} {\bibfnamefont {Y.}~\bibnamefont
  {Asaka}}, \bibinfo {author} {\bibfnamefont {T.}~\bibnamefont {Kikuchi}},\
  and\ \bibinfo {author} {\bibfnamefont {Y.}~\bibnamefont {Fuseya}},\
  }\bibfield  {title} {\bibinfo {title} {Long-range permeation of wave function
  and superficial surface state due to strong quantum size effects in
  topological bi/bisb heterojunctions},\ }\href
  {https://doi.org/10.1103/PhysRevB.106.245303} {\bibfield  {journal} {\bibinfo
   {journal} {Phys. Rev. B}\ }\textbf {\bibinfo {volume} {106}},\ \bibinfo
  {pages} {245303} (\bibinfo {year} {2022})}\BibitemShut {NoStop}%
\bibitem [{\citenamefont {Davis}\ \emph {et~al.}(1992)\citenamefont {Davis},
  \citenamefont {Hannon}, \citenamefont {Ray},\ and\ \citenamefont
  {Plummer}}]{Davis1992}%
  \BibitemOpen
  \bibfield  {author} {\bibinfo {author} {\bibfnamefont {H.~L.}\ \bibnamefont
  {Davis}}, \bibinfo {author} {\bibfnamefont {J.~B.}\ \bibnamefont {Hannon}},
  \bibinfo {author} {\bibfnamefont {K.~B.}\ \bibnamefont {Ray}},\ and\ \bibinfo
  {author} {\bibfnamefont {E.~W.}\ \bibnamefont {Plummer}},\ }\bibfield
  {title} {\bibinfo {title} {Anomalous interplanar expansion at the (0001)
  surface of be},\ }\href {https://doi.org/10.1103/PhysRevLett.68.2632}
  {\bibfield  {journal} {\bibinfo  {journal} {Phys. Rev. Lett.}\ }\textbf
  {\bibinfo {volume} {68}},\ \bibinfo {pages} {2632} (\bibinfo {year}
  {1992})}\BibitemShut {NoStop}%
\bibitem [{\citenamefont {te~Velde}\ and\ \citenamefont
  {Baerends}(1991)}]{Velde1991}%
  \BibitemOpen
  \bibfield  {author} {\bibinfo {author} {\bibfnamefont {G.}~\bibnamefont
  {te~Velde}}\ and\ \bibinfo {author} {\bibfnamefont {E.~J.}\ \bibnamefont
  {Baerends}},\ }\bibfield  {title} {\bibinfo {title} {Precise
  density-functional method for periodic structures},\ }\href
  {https://doi.org/10.1103/PhysRevB.44.7888} {\bibfield  {journal} {\bibinfo
  {journal} {Phys. Rev. B}\ }\textbf {\bibinfo {volume} {44}},\ \bibinfo
  {pages} {7888} (\bibinfo {year} {1991})}\BibitemShut {NoStop}%
\bibitem [{BAN()}]{BAND}%
  \BibitemOpen
  \href@noop {} {}\bibinfo {note} {BAND 2021.1 (SCM, Theoretical Chemistry,
  Vrije Universiteit, Amsterdam, The Netherlands), https://
  www.scm.com/.}\BibitemShut {Stop}%
\bibitem [{\citenamefont {Bitzek}\ \emph {et~al.}(2006)\citenamefont {Bitzek},
  \citenamefont {Koskinen}, \citenamefont {G\"ahler}, \citenamefont {Moseler},\
  and\ \citenamefont {Gumbsch}}]{Bitzek2006}%
  \BibitemOpen
  \bibfield  {author} {\bibinfo {author} {\bibfnamefont {E.}~\bibnamefont
  {Bitzek}}, \bibinfo {author} {\bibfnamefont {P.}~\bibnamefont {Koskinen}},
  \bibinfo {author} {\bibfnamefont {F.}~\bibnamefont {G\"ahler}}, \bibinfo
  {author} {\bibfnamefont {M.}~\bibnamefont {Moseler}},\ and\ \bibinfo {author}
  {\bibfnamefont {P.}~\bibnamefont {Gumbsch}},\ }\bibfield  {title} {\bibinfo
  {title} {Structural relaxation made simple},\ }\href
  {https://doi.org/10.1103/PhysRevLett.97.170201} {\bibfield  {journal}
  {\bibinfo  {journal} {Phys. Rev. Lett.}\ }\textbf {\bibinfo {volume} {97}},\
  \bibinfo {pages} {170201} (\bibinfo {year} {2006})}\BibitemShut {NoStop}%
\bibitem [{\citenamefont {Perdew}\ \emph {et~al.}(1996)\citenamefont {Perdew},
  \citenamefont {Burke},\ and\ \citenamefont {Ernzerhof}}]{Perdew1996}%
  \BibitemOpen
  \bibfield  {author} {\bibinfo {author} {\bibfnamefont {J.~P.}\ \bibnamefont
  {Perdew}}, \bibinfo {author} {\bibfnamefont {K.}~\bibnamefont {Burke}},\ and\
  \bibinfo {author} {\bibfnamefont {M.}~\bibnamefont {Ernzerhof}},\ }\bibfield
  {title} {\bibinfo {title} {Generalized gradient approximation made simple},\
  }\href {https://doi.org/10.1103/PhysRevLett.77.3865} {\bibfield  {journal}
  {\bibinfo  {journal} {Phys. Rev. Lett.}\ }\textbf {\bibinfo {volume} {77}},\
  \bibinfo {pages} {3865} (\bibinfo {year} {1996})}\BibitemShut {NoStop}%
\bibitem [{\citenamefont {van Lenthe}\ \emph {et~al.}(1999)\citenamefont {van
  Lenthe}, \citenamefont {Ehlers},\ and\ \citenamefont
  {Baerends}}]{Lenthe1999}%
  \BibitemOpen
  \bibfield  {author} {\bibinfo {author} {\bibfnamefont {E.}~\bibnamefont {van
  Lenthe}}, \bibinfo {author} {\bibfnamefont {A.}~\bibnamefont {Ehlers}},\ and\
  \bibinfo {author} {\bibfnamefont {E.-J.}\ \bibnamefont {Baerends}},\
  }\bibfield  {title} {\bibinfo {title} {Geometry optimizations in the zero
  order regular approximation for relativistic effects},\ }\href
  {https://doi.org/10.1063/1.478813} {\bibfield  {journal} {\bibinfo  {journal}
  {The Journal of Chemical Physics}\ }\textbf {\bibinfo {volume} {110}},\
  \bibinfo {pages} {8943} (\bibinfo {year} {1999})},\ \Eprint
  {https://arxiv.org/abs/https://pubs.aip.org/aip/jcp/article-pdf/110/18/8943/19297774/8943\_1\_online.pdf}
  {https://pubs.aip.org/aip/jcp/article-pdf/110/18/8943/19297774/8943\_1\_online.pdf}
  \BibitemShut {NoStop}%
\bibitem [{\citenamefont {Koroteev}\ \emph {et~al.}(2008)\citenamefont
  {Koroteev}, \citenamefont {Bihlmayer}, \citenamefont {Chulkov},\ and\
  \citenamefont {Bl\"ugel}}]{Koroteev2008}%
  \BibitemOpen
  \bibfield  {author} {\bibinfo {author} {\bibfnamefont {Y.~M.}\ \bibnamefont
  {Koroteev}}, \bibinfo {author} {\bibfnamefont {G.}~\bibnamefont {Bihlmayer}},
  \bibinfo {author} {\bibfnamefont {E.~V.}\ \bibnamefont {Chulkov}},\ and\
  \bibinfo {author} {\bibfnamefont {S.}~\bibnamefont {Bl\"ugel}},\ }\bibfield
  {title} {\bibinfo {title} {First-principles investigation of structural and
  electronic properties of ultrathin bi films},\ }\href
  {https://doi.org/10.1103/PhysRevB.77.045428} {\bibfield  {journal} {\bibinfo
  {journal} {Phys. Rev. B}\ }\textbf {\bibinfo {volume} {77}},\ \bibinfo
  {pages} {045428} (\bibinfo {year} {2008})}\BibitemShut {NoStop}%
\bibitem [{\citenamefont {Liu}\ and\ \citenamefont {Allen}(1995)}]{Liu1995}%
  \BibitemOpen
  \bibfield  {author} {\bibinfo {author} {\bibfnamefont {Y.}~\bibnamefont
  {Liu}}\ and\ \bibinfo {author} {\bibfnamefont {R.~E.}\ \bibnamefont
  {Allen}},\ }\bibfield  {title} {\bibinfo {title} {Electronic structure of the
  semimetals bi and sb},\ }\href@noop {} {\bibfield  {journal} {\bibinfo
  {journal} {Phys. Rev. B}\ }\textbf {\bibinfo {volume} {52}},\ \bibinfo
  {pages} {1566} (\bibinfo {year} {1995})}\BibitemShut {NoStop}%
\bibitem [{\citenamefont {Ast}\ and\ \citenamefont {H\"ochst}(2003)}]{Ast2003}%
  \BibitemOpen
  \bibfield  {author} {\bibinfo {author} {\bibfnamefont {C.~R.}\ \bibnamefont
  {Ast}}\ and\ \bibinfo {author} {\bibfnamefont {H.}~\bibnamefont {H\"ochst}},\
  }\bibfield  {title} {\bibinfo {title} {Electronic structure of a bismuth
  bilayer},\ }\href {https://doi.org/10.1103/PhysRevB.67.113102} {\bibfield
  {journal} {\bibinfo  {journal} {Phys. Rev. B}\ }\textbf {\bibinfo {volume}
  {67}},\ \bibinfo {pages} {113102} (\bibinfo {year} {2003})}\BibitemShut
  {NoStop}%
\bibitem [{\citenamefont {Koroteev}\ \emph {et~al.}(2004)\citenamefont
  {Koroteev}, \citenamefont {Bihlmayer}, \citenamefont {Gayone}, \citenamefont
  {Chulkov}, \citenamefont {Bl\"ugel}, \citenamefont {Echenique},\ and\
  \citenamefont {Hofmann}}]{Koroteev2004}%
  \BibitemOpen
  \bibfield  {author} {\bibinfo {author} {\bibfnamefont {Y.~M.}\ \bibnamefont
  {Koroteev}}, \bibinfo {author} {\bibfnamefont {G.}~\bibnamefont {Bihlmayer}},
  \bibinfo {author} {\bibfnamefont {J.~E.}\ \bibnamefont {Gayone}}, \bibinfo
  {author} {\bibfnamefont {E.~V.}\ \bibnamefont {Chulkov}}, \bibinfo {author}
  {\bibfnamefont {S.}~\bibnamefont {Bl\"ugel}}, \bibinfo {author}
  {\bibfnamefont {P.~M.}\ \bibnamefont {Echenique}},\ and\ \bibinfo {author}
  {\bibfnamefont {P.}~\bibnamefont {Hofmann}},\ }\bibfield  {title} {\bibinfo
  {title} {Strong spin-orbit splitting on bi surfaces},\ }\href
  {https://doi.org/10.1103/PhysRevLett.93.046403} {\bibfield  {journal}
  {\bibinfo  {journal} {Phys. Rev. Lett.}\ }\textbf {\bibinfo {volume} {93}},\
  \bibinfo {pages} {046403} (\bibinfo {year} {2004})}\BibitemShut {NoStop}%
\bibitem [{\citenamefont {Hirahara}\ \emph {et~al.}(2006)\citenamefont
  {Hirahara}, \citenamefont {Nagao}, \citenamefont {Matsuda}, \citenamefont
  {Bihlmayer}, \citenamefont {Chulkov}, \citenamefont {Koroteev}, \citenamefont
  {Echenique}, \citenamefont {Saito},\ and\ \citenamefont
  {Hasegawa}}]{Hirahara2006}%
  \BibitemOpen
  \bibfield  {author} {\bibinfo {author} {\bibfnamefont {T.}~\bibnamefont
  {Hirahara}}, \bibinfo {author} {\bibfnamefont {T.}~\bibnamefont {Nagao}},
  \bibinfo {author} {\bibfnamefont {I.}~\bibnamefont {Matsuda}}, \bibinfo
  {author} {\bibfnamefont {G.}~\bibnamefont {Bihlmayer}}, \bibinfo {author}
  {\bibfnamefont {E.~V.}\ \bibnamefont {Chulkov}}, \bibinfo {author}
  {\bibfnamefont {Y.~M.}\ \bibnamefont {Koroteev}}, \bibinfo {author}
  {\bibfnamefont {P.~M.}\ \bibnamefont {Echenique}}, \bibinfo {author}
  {\bibfnamefont {M.}~\bibnamefont {Saito}},\ and\ \bibinfo {author}
  {\bibfnamefont {S.}~\bibnamefont {Hasegawa}},\ }\bibfield  {title} {\bibinfo
  {title} {Role of spin-orbit coupling and hybridization effects in the
  electronic structure of ultrathin bi films},\ }\href
  {https://doi.org/10.1103/PhysRevLett.97.146803} {\bibfield  {journal}
  {\bibinfo  {journal} {Phys. Rev. Lett.}\ }\textbf {\bibinfo {volume} {97}},\
  \bibinfo {pages} {146803} (\bibinfo {year} {2006})}\BibitemShut {NoStop}%
\bibitem [{\citenamefont {Oura}\ \emph {et~al.}(2003)\citenamefont {Oura},
  \citenamefont {Katayama}, \citenamefont {Zotov}, \citenamefont {Lifshits},\
  and\ \citenamefont {Saranin}}]{Oura_book}%
  \BibitemOpen
  \bibfield  {author} {\bibinfo {author} {\bibfnamefont {K.}~\bibnamefont
  {Oura}}, \bibinfo {author} {\bibfnamefont {M.}~\bibnamefont {Katayama}},
  \bibinfo {author} {\bibfnamefont {A.~V.}\ \bibnamefont {Zotov}}, \bibinfo
  {author} {\bibfnamefont {V.~G.}\ \bibnamefont {Lifshits}},\ and\ \bibinfo
  {author} {\bibfnamefont {A.~A.}\ \bibnamefont {Saranin}},\ }\href@noop {}
  {\emph {\bibinfo {title} {Surface Science: An Introduction}}}\ (\bibinfo
  {publisher} {Springer Berlin, Heidelberg},\ \bibinfo {year}
  {2003})\BibitemShut {NoStop}%
\bibitem [{\citenamefont {Hsieh}\ \emph {et~al.}(2012)\citenamefont {Hsieh},
  \citenamefont {Lin}, \citenamefont {Liu}, \citenamefont {Duan}, \citenamefont
  {Bansil},\ and\ \citenamefont {Fu}}]{Hsieh2012}%
  \BibitemOpen
  \bibfield  {author} {\bibinfo {author} {\bibfnamefont {T.~H.}\ \bibnamefont
  {Hsieh}}, \bibinfo {author} {\bibfnamefont {H.}~\bibnamefont {Lin}}, \bibinfo
  {author} {\bibfnamefont {J.}~\bibnamefont {Liu}}, \bibinfo {author}
  {\bibfnamefont {W.}~\bibnamefont {Duan}}, \bibinfo {author} {\bibfnamefont
  {A.}~\bibnamefont {Bansil}},\ and\ \bibinfo {author} {\bibfnamefont
  {L.}~\bibnamefont {Fu}},\ }\bibfield  {title} {\bibinfo {title} {Topological
  crystalline insulators in the snte material class},\ }\href
  {https://doi.org/10.1038/ncomms1969} {\bibfield  {journal} {\bibinfo
  {journal} {Nature Communications}\ }\textbf {\bibinfo {volume} {3}},\
  \bibinfo {pages} {982} (\bibinfo {year} {2012})}\BibitemShut {NoStop}%
\end{thebibliography}%

\end{document}